\documentclass[apjpt4]{aastex}

\def\url#1{{\ttfamily\def\/{/\discretionary{}{}{}}#1}}

\def\mathnew{\mathsurround=0pt}
\def\simov#1#2{\lower .5pt\vbox{\baselineskip0pt
    \lineskip-.5pt\ialign{$\mathnew#1\hfil##\hfil$\crcr#2\crcr\sim\crcr}}}  
\def\simgreat{\mathrel{\mathpalette\simov >}}

\def\'#1{\ifx#1i{\accent"13\i}\else{\accent"13#1}\fi}

\shorttitle{DM and DE vs.~CMB data}
\shortauthors{Mainini et al.}
\begin{document}
\title{Dark Matter and Dark Energy from a single scalar field and CMB data.}

\author{Roberto Mainini, Loris P.L. Colombo \& Silvio A. Bonometto}

\affil{Physics Department G. Occhialini, Universit\`a degli Studi di
Milano--Bicocca, Piazza della Scienza 3, I20126 Milano (Italy) 
\& I.N.F.N., Sezione di Milano}

\begin{abstract}
Axions are likely to be the Dark Matter (DM) that cosmological data
require. They arise in the Peccei--Quinn solution of the strong--$CP$
problem. In a previous work we showed that their model has a simple
and natural generalization which yields also Dark Energy (DE), in fair
proportions, without tuning any parameter: DM and DE arise from a
single scalar field and are weakly coupled in the present era.  In
this paper we extend the analysis of this {\it dual--axion} cosmology
and fit it to WMAP data, by using a Markov chain technique.  We
find that $\Lambda$CDM, dynamical DE with a SUGRA potential, DE with a
SUGRA potential and a constant DE--DM coupling, {\it as well as the
dual--axion model} with a SUGRA potential, fit data with a similar
accuracy. The best--fit parameters are however fairly different,
although consistency is mostly recovered at the 2--$\sigma$ level. A
peculiarity of the dual--axion model with SUGRA potential is to cause
more stringent constraints on most parameters and to favor high
values of the Hubble parameter.
\end{abstract}


%=====================
\section{Introduction}
Models with density parameters $\Omega_{o,de} \simeq 0.7$,
$\Omega_{o,m} \simeq 0.3$, $\Omega_{o,b} \simeq 0.04$ (for dark
energy, whole non--relativistic matter and baryons, respectively),
Hubble parameter $h\simeq0.7$ (in units of 100 km/s/Mpc) and primeval
spectral index $n_s\simeq1$ fit most cosmological data, including
Cosmic Microwave Background (CMB) anisotropies, large scale structure,
as well as data on SNIa (Tegmark et al. 2001, De Bernardis et
al. 2000, Hanany et al. 2000, Halverson et al. 2001, Spergel et
al. 2003, Percival et al. 2002, Efstathiou et al. 2002, Riess et
al. 1988, Perlmutter et al.  1988). The success of such $\Lambda$CDM
models, also dubbed cosmic concordance models, does not hide their
uneasiness. The parameters of standard CDM are still to be increased
by one, in order to tune Dark Energy (DE).  Furthermore, if DE is
ascribed to vacuum, this turns out to be quite a fine tuning.

This conceptual problem was eased by dynamical DE models (Wetterich 1988,
1995; Ratra \& Pee\-bles 1988, RP hereafter). They postulate the
existence of an {\it ad--hoc} scalar field, self--interacting through
a suitable effective potential, which depends on a further
parameter. In RP and SUGRA (see below) models, this is an energy scale
$\Lambda$ (or an exponent $\alpha$).

Within the frame of dynamical DE models, Mainini \& Bonometto (2004,
MB hereafter) tried to take a step forward. Instead of invoking an
{\it ad--hoc} interaction, they refer to the field introduced by
Peccei \& Quinn (1977, PQ hereafter) to solve the strong--CP
problem. If suitably tuned, such scheme was already shown to yield DM
(Preskill, Wise \& Wilczek 1983, Abbott \& Sikivie 1983, Dine \&
Fischler 1983). MB slightly modify the PQ scheme, replacing the
Nambu--Goldstone (NG) potential introduced {\it ad--hoc}, by a
potential admitting a tracker solution. This scheme solves the
strong--CP problem even more efficiently than the original PQ model.
The $\Lambda$ parameter of the tracker potential takes the place of
the PQ energy scale, $F_{PQ}$. Fixing it in the range solving the
strong--CP problem yields DM and DE in fair proportions. Here, we
shall call this cosmology {\it dual--axion} model. This model has
several advantages both in respect to $\Lambda$CDM and ordinary
dynamical DE: (i) it requires no fine tuning; (ii) it adds no
parameter to the standard PQ scheme, which yields just DM; (iii) it
introduces no field or interaction, besides those required by particle
physics. This scheme, however, leads to predictions (slightly)
different from $\Lambda$CDM, for a number of observables. In
principle, therefore, it can be falsified by data.

The essential peculiarity of the model is that it predicts a coupling
between DM and DE. Coupled DE models were introduced by a number of
authors (see, e.g., Amendola 2000, 2003; Gasperini, Piazza \&
Veneziano 2002; Perrotta \& Baccigalupi 2002) as no direct evidence
exists that DM particles follow geodesics. Most such models, however,
introduce a further coupling parameter $\beta$. Its tuning fixes
DM--DE coupling within an acceptable range. For instance, Amendola \&
Quercellini (2003) give limits on $\beta$ deduced from a fit to WMAP
data (Spergel et al 2003); Macci\`o et al. (2004) restricted $\beta$
even more, by studying the halo profiles, produced in N--body
simulations with coupled DE.

At variance from these pictures, the dual--axion scheme has no extra
coupling parameter. The strength of the coupling is set by theory and,
if this conflicts with data, the whole scheme is falsified.  The only
degree of freedom still allowed is the choice of the tracker
potential. This freedom exists for any dynamical DE model. Also PQ
exploited it by choosing a NG potential. For the sake of
definiteness, up to now, the dual--axion scheme has been explored just
in association with a SUGRA potential. MB showed that, in this case,
the dual axion scheme predicts a fair growth of density fluctuations,
so granting a viable picture for the Large Scale Structure.

The recent detailed WMAP data on CMB anisotropies allow to submit the
dual--axion model to further stringent tests, by comparing it with
other cosmologies as $\Lambda$CDM, standard and interacting dynamical
DE. This is done here by using a multiparameter Markov chain technique
\citep[see, e.g., Kosowski et al. 2003;][]{christensen,knox,lewis}.
The results of this numerical approach will then be discussed and
understood on physical bases.  None of the above models however
performs neatly better than the others.  Apparently, the best fit is
obtained by dynamical DE based on a SUGRA potential, but its success
is strictly marginal. The analysis of these models against CMB data is
the main aim of this paper.

The plan of the paper is as follows: In Section 2 we summarize the
particle physics background to the dual axion model. In particular,
starting from Section 2.3, we restrict our analysis to a particular
form of DE potential, the SUGRA potential; using a different potential
might change some quantitative results. In Section 3 we describe the
technique used to compare different models with CMB data and present
(subsection 3.2) the results of this comparison. Section 4 is devoted
to a final discussion of the results.

\section{A single scalar field to account for DM and DE}
Let us first remind that the strong $CP$ problem arises from the
existence in QCD (quantum--chromo--dynamics) of multiple vacuum
states.  The set of the gauge transformations $\Omega(x_i)$ that join
vacuum configurations can be subdivided in classes ${\Omega_n(x_i)}$,
characterized by an integer $n$ (Jackiw \& Rebbi 1976), setting their
different asymptotic behaviors. Within each class, transformations can
be distorted into each other with continuity, while this is impossible
if they pertain to different classes.

Accordingly, in classical field theory there is no communication
between different--$n$ gauge sectors. In quantum field theory,
instead, tunneling is possible thanks to instanton effects, so that
any vacuum state is a superposition of the vacua $|0_n\rangle$ (of the
$n$th sector), of the kind $|0_\theta\rangle = \sum_n |0_n\rangle
\exp(in\theta)$.

The effects of varying the $\theta$--vacuum can be recast into 
variations of a non--perturbative term 
\begin{equation}
{\cal L}_\theta = {\alpha_s \over 2\pi} \theta \, G \cdot {\tilde G}
\label{eq:n1}
\end{equation}
($\alpha_s$: strong coupling constant, $G$ and $ {\tilde G}$: gluon
field tensor and its dual) in the QCD Lagrangian density.
However, chiral transformations also change the vacuum angle, so that
the $\theta$--parameter receives
another contribution, arising from the EW (electro-weak) sector, when the
quark mass matrix $\cal M$ is diagonalized, becoming
\begin{equation}
\theta_{eff} = \theta + Arg ~det ~{\cal M}~.
\end{equation} 
The Lagrangian term (\ref{eq:n1}) can be reset in the form of a 4--divergence
and causes no change of the equations of motion. It however violates
$CP$ and, among various effects, yields a neutron electric moment $d_n
\simeq 5 \cdot 10^{-16} \theta_{eff} {\rm ~e~cm}$, conflicting with
the experimental limit $d_n \lesssim 10^{-25} {\rm ~e~cm}$, unless
$\theta_{eff}   \lesssim 10^{-10}$. 

The point is that the two contributions to $\theta_{eff}$ are
uncorrelated, so that there is no reason why their sum should be so
small.

PQ succeed in suppressing this term by imposing an additional global
chiral symmetry $U(1)_{PQ}$, spontaneously broken at a suitable scale
$F_{PQ}$. The axion field is suitably coupled to the quark sector.
The details of this coupling depend on the model and may require the
introduction of an {\it ad--hoc} heavy quark (Kim 1979, Shifman et
al. 1979, see also Dine et al. 1981, Zhitnisky 1980).  The $U(1)_{PQ}$
symmetry suffers from a chiral anomaly, so the axion acquires a tiny
mass because of non-perturbative effects, whose size has a rapid
increase around the quark-hadron transition scale $\Lambda_{QCD}$.
The anomaly manifests itself when a chiral $U(1)_{PQ}$ transformation
is performed on the axion field, giving rise to a term of the same
form of (\ref{eq:n1}), which provides a potential for the axion field.

As a result, $\theta$ is effectively replaced by the dynamical axion
field. Its oscillations about the potential minimum yield axions. This
mechanism works independently of the scale $F_{PQ}$. Limits on it
arise from astrophysics and cosmology, requiring that $10^{10}
GeV \lesssim F_{PQ} \lesssim 10^{12}GeV$; in turn, this yields an
axion mass which lays today in the interval $10^{-6} eV \lesssim m_A
\lesssim 10^{-3} eV$.

More in detail, in most axion models, the PQ symmetry breaking occurs
when a complex scalar field $\Phi = \phi e^{i\theta}/\sqrt{2}$, falling
into one of the degenerated minima of a NG potential
\begin{equation}
V(\Phi) = \lambda [|\phi|^2 - F_{PQ}^2]^2 ~,
\label{eq:n2}
\end{equation}
develops a vacuum expectation value $\langle \phi \rangle= F_{PQ}$. 

The $CP$-violating term, arising around quark-hadron transition when
$\bar q q$ condensates break the chiral symmetry, reads
\begin{equation}
V_1 = \left[\sum_q \langle 0(T)| {\bar q} q |0(T) \rangle m_q \right] 
~(1 - \cos \theta)
\label{eq:n3}
\end{equation}
($\sum_q$ extends over all quarks), so that $\theta$ is no longer
arbitrary, but shall be ruled by a suitable equation of motion. The
term in square brackets, at $T \simeq 0$, approaches $m_\pi^2 f_\pi^2$
($m_\pi$ and $f_\pi$: $\pi$--meson mass and decay constant).  In this
limit, for $\theta \ll 1$ and using $A=\theta F_{PQ}$ as axion field,
eq.~(\ref{eq:n3}) reads:
\begin{equation}
V_1 \simeq  {1\over 2} q^2(m_q) m_\pi^2 f_\pi^2 {A^2 \over F_{PQ}^2}~;
\label{eq:n4}
\end{equation}
here $q(m_q)$ is a function of the quark masses $q_i$; in the limit
of 2 light quarks ($u$ and $d$),  $q = \sqrt{m_u/m_d}(1+m_u/m_d)^{-1}$.
The $A$ field bears the right dimensions but, here below, will no
longer be used, and the axion degrees of freedom will be described
through $\theta$ itself. Eq.~(\ref{eq:n4}), however, shows that, when
$\langle \bar q q \rangle$ is no longer zero (since $T \lesssim
\Lambda_{QCD}$), the axion mass decreases with temperature approaching
the constant value $m_{A}={m_\pi f_\pi q(m_q) / F_{PQ}}$ for $T \ll
\Lambda_{QCD}$.

Accordingly, the equation of motion, in the small $\theta$ limit, reads
\begin{equation} 
\ddot \theta + 2{\dot a \over a} 
\dot \theta +  a^2 m_A^2  \theta = 0~,
\end{equation} 
(here $a$ is the scale factor and dots yield differentiation with
respect to conformal time, see next Section), so that the axion field
undergoes (nearly) harmonic oscillations, as soon as $m_A$ exceeds
the expansion rate; then, his mean pressure vanishes (Dine \&
Fischler 1983) leaving axion as a viable candidate for cold DM.

\vglue 0.05truecm

MB replace the NG potential in eq.~(\ref{eq:n2}) by a potential
$V(\Phi)$ admitting a tracker solution (Wetterich 1988,1995, RP 1988,
Ferreira \& Joyce 1998, Brax \& Martin 1999, 2001, Brax, Martin \&
Riazuelo 2000). The field $\Phi$ is complex and $V(\Phi)$ is $U(1)$
invariant, but there is no transition to a constant value $F_{PQ}$,
which is replaced by the modulus $\phi$ itself, slowly evolving over
cosmological times. At a suitable early time, quantum dynamics starts
to be fairly accounted by the potential $V$; soon after, $\phi$
settles on the tracker solution in almost any horizon, however
breaking the $U(1)$ symmetry, by the values assumed by $\theta$, in
different horizons. Later on, when chiral symmetry breaks, dynamics
becomes relevant also for the $\theta$ degree of freedom, as in the PQ
case. At variance from it, however, this happens while $\phi$
continues its slow evolution, down to the present epoch, when it
accounts for DE. Owing to the $\phi$ evolution, however, the axion
mass evolves, over cosmological times, also for $T \ll \Lambda_{QCD}$
(see below).

The $\Phi$ field, therefore, besides of providing DM through its phase
$\theta$, whose dynamics solves the strong $CP$ problem, also accounts
for DE through its modulus $\phi$.

This scheme holds for any DE potential admitting tracker solutions.
To be more specific, MB use a SUGRA potential \citep[Brax, Martin
\& Riazuelo 2000;][]{BraxMartin99, BraxMartin01} finding that, at the
quark--hadron transition, $\phi$ can be naturally led to have values
$\sim F_{PQ}$, increasing up to $\sim m_p = G^{-1/2}$ (the Planck
mass), when approaching today. The only free parameter is the energy
scale in the SUGRA potential, that must be $\sim 10^{10}$GeV.  With
this choice, $\theta$ is driven to values even smaller than in the PQ
case, so that $CP$ is apparently conserved in strong interactions,
while $\Omega_{o,m}$, $\Omega_{o,de}$ (the DE density parameter) and
$\Omega_{o,b}$ take fair values.

\subsection{Lagrangian theory}
In the dual--axion model we start from the Lagrangian
\begin{equation}
{\cal L} =  \sqrt{-g} \{ g_{\mu\nu} 
\partial_\mu \Phi \partial_\nu \Phi   - V(\phi) \} ~,
\end{equation}
which can be rewritten in terms of $\phi$ and $\theta$, adding
also the term breaking the $U(1)$ symmetry. Then it reads:
\begin{equation}
%\begin{eqnarray}
{\cal L} = \sqrt{-g} \left\{ {1 \over 2} 
g_{\mu\nu} [\partial_\mu \phi \partial_\nu \phi 
+ \phi^2  \partial_\mu \theta \partial_\nu \theta] 
%\nonumber \\
- V(\phi) -m^2(T,\phi) \phi^2 (1 - \cos \theta) 
\right\}  ~.
\label{eq:m1}
%\end{eqnarray}	
\end{equation}
Here $g_{\mu\nu}$ is the metric tensor. We shall assume that $ds^2 =
g_{\mu\nu} dx^\mu dx^\nu = a^2 (d\tau^2 - \eta_{ij}dx_idx_j)$, so that
$a$ is the scale factor, $\tau$ is the conformal time; Greek (Latin)
indexes run from 0 to 3 (1 to 3); dots indicate differentiation in
respect to $\tau$. The mass behavior for $T \sim \Lambda_{QCD}$ will
be detailed in Section 2.2. The equations of motion, for the $\phi$
and $\theta$ degrees of freedom, read
\begin{equation}
\ddot \theta + 2\left({\dot a \over a}+{\dot \phi \over \phi}\right) 
\dot \theta + m^2 a^2 \sin \theta = 0~,
\label{eq:m3}
\end{equation}
\begin{equation}
\ddot \phi + 2 {\dot a \over a}  \dot \phi + 
a^2  V'(\phi) = \phi\, \dot \theta^2 ~.
\label{eq:m4}
\end{equation}
(Notice that $m^2(T,\phi) \phi^2$ is $\phi$ independent, see below.)
In what follows, the former equation will be always considered when
$\sin \theta \simeq \theta$. In particular, taking into account the
condition $\theta \ll 1$, the expressions for the energy densities
$\rho_{\theta,\phi} = \rho_{\theta,\phi;kin} + \rho_{\theta, \phi;
pot}$ and the pressures $p_{\theta,\phi} = \rho_{\theta,\phi;kin} -
\rho_{\theta, \phi;pot}$ are obtainable by combining the terms
$$	
~~~~~ \rho_{\theta,kin} = {\phi^2 \over 2 a^2} \dot \theta^2~,~~
\rho_{\theta,pot} = m^2(T,\phi) \phi^2 (1-\cos \theta) \simeq
{m^2(T,\phi) \over 2} \phi^2 \theta^2~,~~ 
$$
\begin{equation}
\\ \rho_{\phi,kin}
= {\dot \phi^2 \over 2 a^2} ~~,~~~~ \rho_{\phi,pot} = V(\phi)~.~~~~~~~
~~~~~~~~~~~~~~~~~~~~~~~~~~~~~~~~~~~
\label{enpre}
\end{equation}	
When $\theta$ undergoes many (nearly) harmonic oscillations within a
Hubble time, $\langle \rho_{\theta,kin} \rangle \simeq \langle
\rho_{\theta,pot} \rangle$ and $\langle p_\theta \rangle$ vanishes
(Dine \& Fischler 1983).  Under such condition, using
eqs.~(\ref{eq:m3}) and (\ref{eq:m4}), it is easy to see that
\begin{equation}
\dot \rho_\theta + 3{\dot a \over a} \rho_\theta = {\dot m \over m}
 \rho_\theta
~,~~ \dot \rho_\phi + 3 {\dot a \over a} (\rho_\phi+p_\phi)
= - {\dot m \over m}  \rho_\theta ~.
\label{eq:m7}
\end{equation}
When $m$ is given by eq.~(15) and (16) here below, $\dot m/m
= -\dot \phi/\phi - 3.8\, \dot T/T$. 
At $T \simeq 0$, instead, it is just  $\dot m / m \simeq -\dot \phi/\phi$.

The $\theta$ and $\phi$ components account for DM and DE,
respectively. Accordingly, in the sequel, the indexes 
$_\theta$, $_\phi$ will be replaced by $_{dm}$, $_{de}$.

\begin{figure}
\plotone{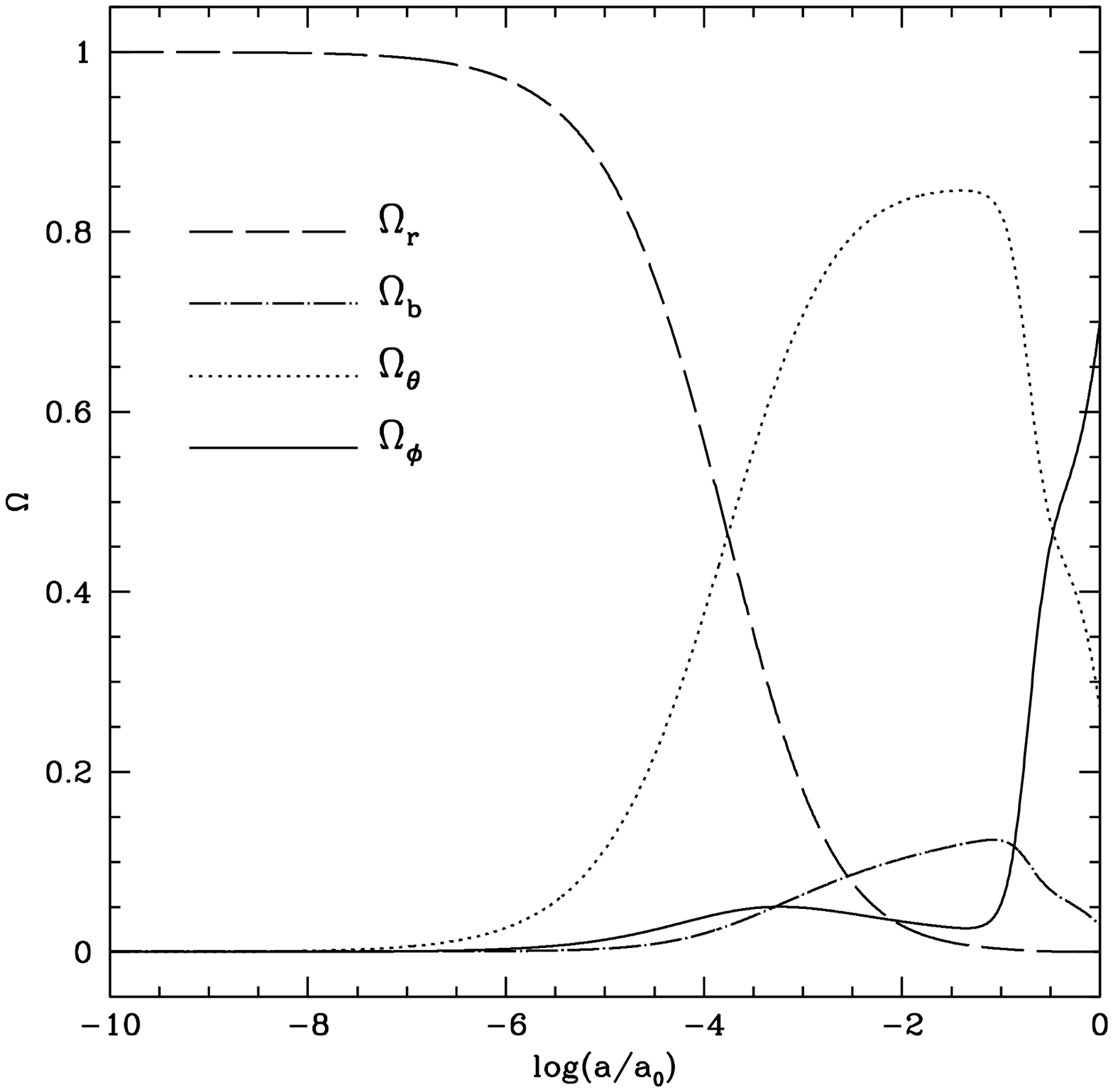}
\epsscale{1.0}
\caption{Density parameters $\Omega_i$ ($i = r,~b,~\theta,~\phi$, {\it
i.e.}  radiation, baryons, DM, DE).}
\label{f02}
\end{figure}

Eqs.~(\ref{eq:m7}) clearly show that an exchange of energy occurs
between DM and DE. From this point of view, the MB model belongs to
the set of coupled models treated by Amendola (2000, 2003). It is
however characterized by a time--dependent coupling.  In fact, in the
small $\theta$ limit, after averaging over cosmological times, the
r.h.s. of eqs.~(\ref{eq:m4}) and (\ref{eq:m7}) read $ C(\phi) \langle
\rho_\theta \rangle a^2$ and $\pm C(\phi) \, \dot\phi\, \langle
\rho_\theta \rangle$, if we set $C(\phi ) = 1/\phi$.  Here $C$ is the
DE--DM coupling introduced by Amendola (2000,~2003), who however
studies extensively only the case $C = \beta (16 \pi / 3
m_p^2)^{1/2}$, with constant $\beta$. In these latter DE models, a
$\phi$--$MDE$ phase takes place after matter--radiation
equivalence. It differs from a matter dominated expansion because of
the contribution of the kinetic part of the DE field to the expansion
source.  A regime of this kind is present also in the dual--axion
model and is shown in Fig.~\ref{f02}, here below. Because of the $\phi$
dependence, however, the DM--DE coupling, in the dual--axion model,
weakens as we approach the present cosmological epoch.

The most stringent limits on $\beta$ are set by non--linear
predictions (Macci\`o et al. 2004) and restrict $\beta$ to values
$\lesssim 0.1$--0.2$\, $, in order to avoid a too high concentration
in DM halos. In turn, this is due to the behavior of the effective
mass of DM particles, in the presence of the $\beta$--coupling. A
preliminary inspection indicates that the behavior expected here is
opposite and that halo concentrations, in average, should be smaller
than in $\Lambda$CDM. This point, however, must be inspected in much
more detail.

Let us also notice that the former eq.~(\ref{eq:m7}) can be
integrated soon, yielding $\rho_{dm} \propto m/a^3$. In particular, 
this law holds at $T \ll \Lambda_{QCD}$.
Accordingly, at late times
\begin{equation}
\rho_{dm} a^3 \phi \simeq {\rm const.},
\label{eq:m8}
\end{equation}
so that the usual behavior $\rho_{dm} \propto a^{-3} $ is modified
by the energy outflow from DM to DE. This modification is stronger
when $\phi$ varies rapidly and is damped when $\phi$ attains a
nearly constant behavior.

%%%%%%%%%%%%%%%%%%%%%%%%%%%%%%%%%%%%%%%%%%%%%%%%%%%%%%%%%%%%%%%%%%%%%
\begin{figure}
\plotone{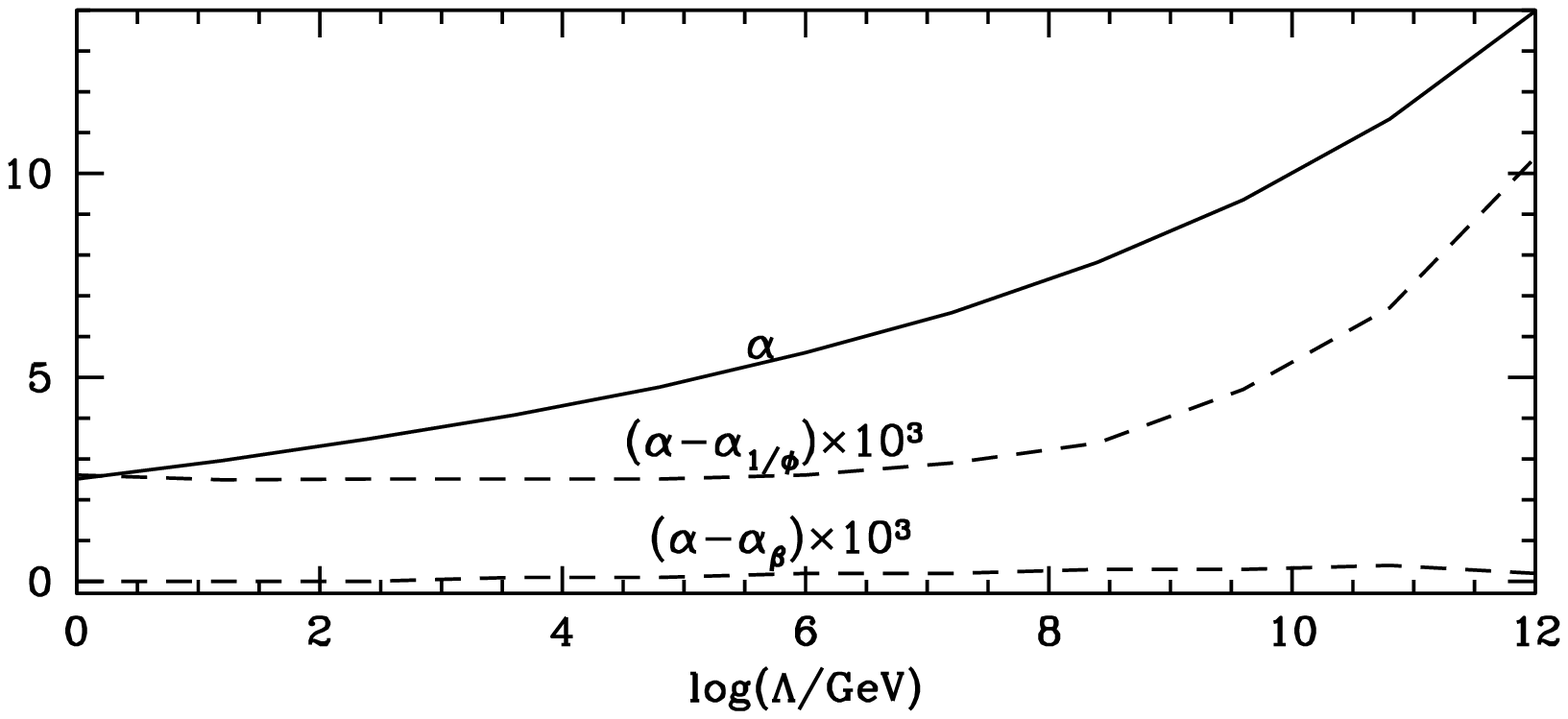}
\epsscale{1.0}
\caption{Values of $\alpha$ corresponding to given $\Lambda$ scales in
SUGRA models with $\Omega_{o,dm} = 0.27$. The tiny differences, just
above noise level, for coupled models ($\alpha_\beta$ and
$\alpha_{1/\phi}$ for constant coupling and $\phi^{-1}$ coupling,
respectively), are also shown.}
\label{fal}
\end{figure}
%%%%%%%%%%%%%%%%%%%%%%%%%%%%%%%%%%%%%%%%%%%%%%%%%%%%%%%%%%%%%%%%%%%%%
\subsection{Axion mass}
\begin{figure}
\plotone{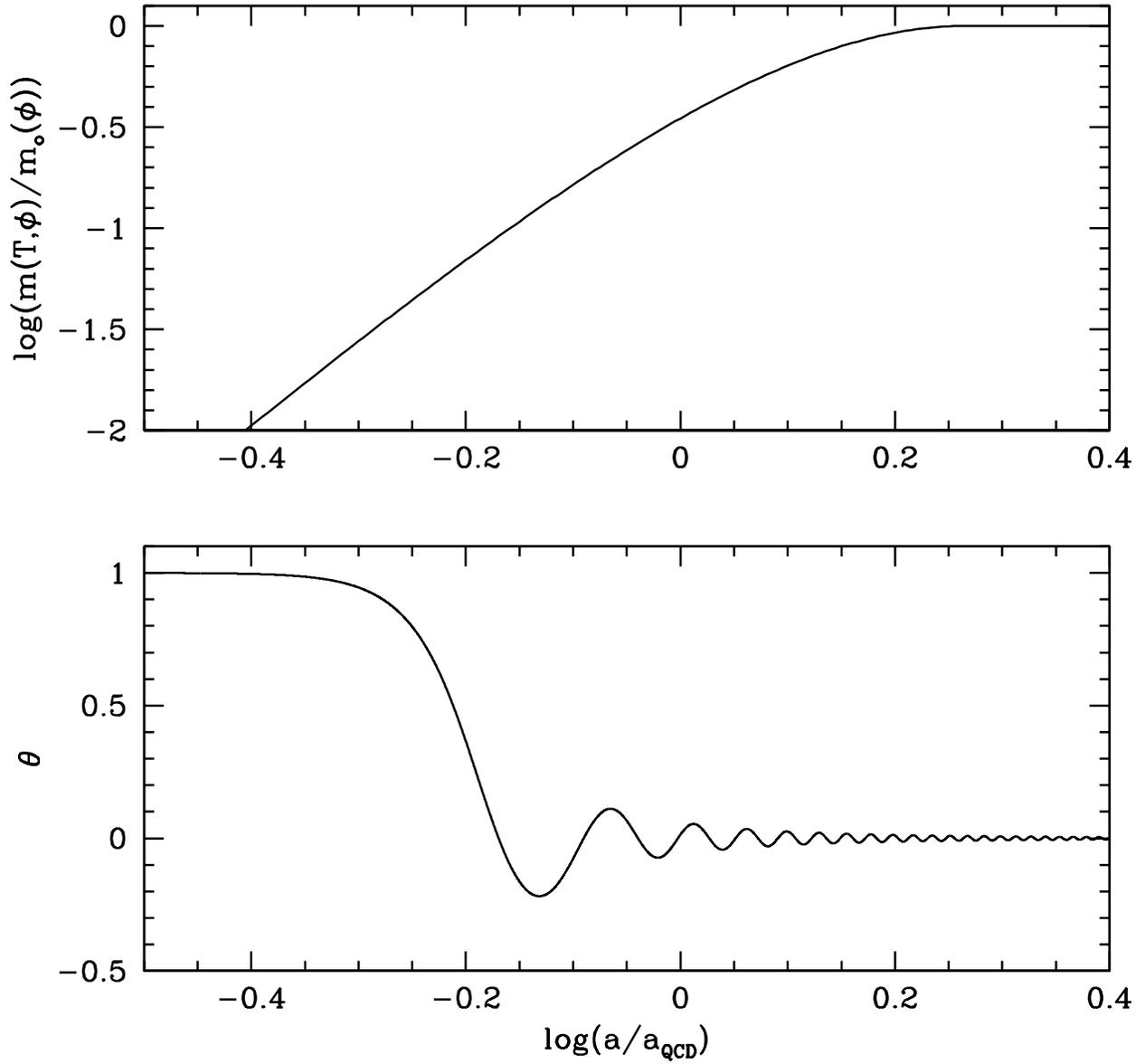}
\epsscale{1.0}
\caption{Interpolated mass behavior and first $\theta$
oscillations. $a_{QCD}$ is the scale when $T=\Lambda_{QCD}$.}
\label{f00}
\end{figure}

According to eq.~(\ref{eq:m3}), the axion field begins to oscillate when:
\begin{equation}
m(T,\phi)a \simeq 2\left({\dot a \over a}+{\dot \phi \over \phi}\right)  ~.
\label{eq:o1}
\end{equation}
In the dual--axion model, just as for PQ, the axion mass rapidly
increases when the chiral symmetry is broken by the formation of the
$\bar q q$ condensate at $T \sim \Lambda_{QCD}$. 
%Around such $T$, therefore, the axion mass grows rapidly. 
In the dual--axion model,
however, 
%a further growth takes place 
the axion mass varies also later on, because of the evolution of
$\phi$, when $m(T,\phi)$ is
%SB  cambiato 130 e m_o a T=0
\begin{equation}
m_o(\phi) = {q(m_q)\, m_\pi f_\pi \over \phi} 
\simeq {\mu_{QCD}^2 \over \phi} ~;
\label{eq:o2}
\end{equation}
with $\mu_{QCD} \simeq 80\,$MeV.  Since $\phi \sim m_p$ today, the
present axion mass $m_a \sim 5 \cdot 10^{-13}$eV.  At high
temperature, according to Gross et al. (1981),
\begin{equation}	
m(T,\phi) \simeq 0.1\, m_o(\phi) \left( \Lambda_{QCD} \over T 
\right)^{3.8}
\label{eq:o3}
\end{equation} 
This expression must be interpolated with eq.~(\ref{eq:o2}), to study
the fluctuation onset for $T \sim \Lambda_{QCD}$.  We report the
results of the solution of the equations in Sec.~2.1, obtained by
assuming
\begin{equation}
m(T,\phi) = \, m_o(\phi) \left(0.1^{^{1/3.8}} \Lambda_{QCD} \over T
\right)^{3.8(1-{a/ a_c})}~~~a<a_c \nonumber 
\label{o4a} 
\end{equation} 
\begin{equation}
m(T,\phi) = m_o(\phi)
~~~~~~ ~~~~\, ~~~~~~~~~~~~~~~~~~~~~~~~~~~~a>a_c \nonumber
\label{o4b} 
\end{equation}
with $a_c = T_o/\Lambda_{QCD}\, 0.1^{1/3.8} = 2.16 \cdot 10^{-12}$;
here $T_o = 2.35 \cdot 10^{-4}\, $eV, $\Lambda_{QCD} = 200\, $MeV$
\simeq 2.5 \, \mu_{QCD}$. The expressions (\ref{o4a}), (\ref{o4b})
assure that $m(T,\phi)$ and its first derivative are continuous and
that the axion mass meets its low--$T$ behavior $m_o(T)$ at $T \simeq
0.55 \, \Lambda_{QCD}$.  The effects of selecting a different value
for $\Lambda_{QCD}$ were studied. Also the impact of the assumed $T$
dependence was considered.  A change of $\Lambda_{QCD}$, by a factor
2, yields a variation of $\Lambda$ in the SUGRA potential (see next
section) by not more than a few percent. Taking a power $\nu \neq 1$
of $(1-a/a_c)$, in the exponent, instead, does not affect continuity
but changes the rapidity by which the high--$T$ regime (\ref{eq:o3})
is met. In Fig.~\ref{f00} we plot both the interpolated mass behavior
and the first $\theta$ oscillations. Changing $\nu$ just slightly
displaces the interpolating curve between $\sim a_{QCD}$ and $\sim 1.5\,
a_{QCD}$, causing a minor phase shift.  Let us finally outline that
eqs.~(\ref{eq:o2}) and (\ref{eq:o3}), as well as the interpolation
(\ref{o4a}),(\ref{o4b}), assure that $m(T,\phi)\phi$ is $\phi
$--independent, as is required to give the equation of motion the form
(\ref{eq:m4}).

%%%%%%%%%%%%%%%%%%%%%%%%%%%%%%%%%%%%%%%%%%%%%%%%%%%%%%%%%%%%%%%%%%%%%
\begin{figure}
\plotone{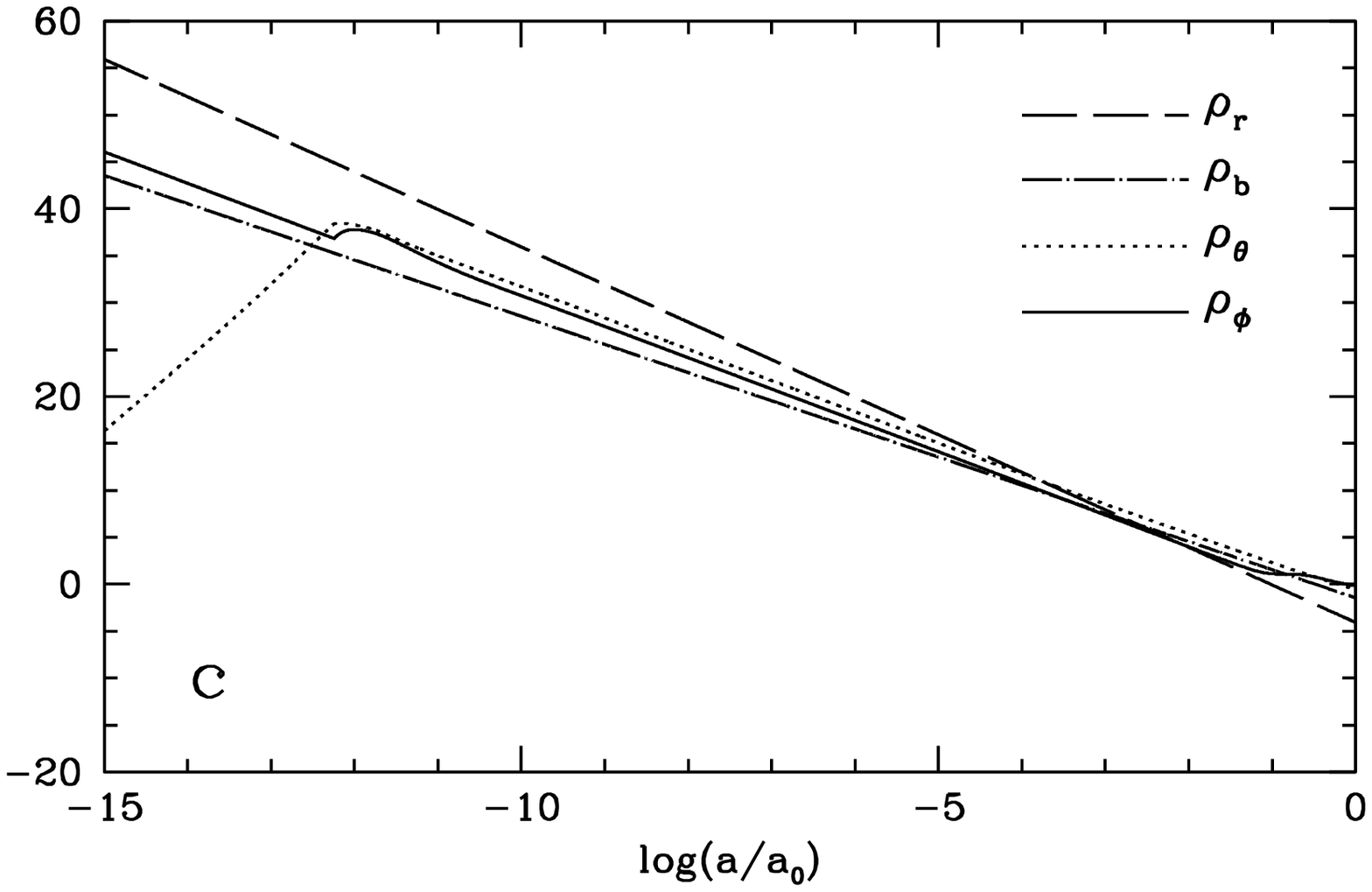}
\epsscale{1.0}
\caption{Behaviors of densities from the tracking regimes down to the
regimes when DE density exceeds radiation ($z \sim 100$), baryons (at
$z \sim 10$) and DM (at $z \sim 3$).}
\label{f01}
\end{figure}
%%%%%%%%%%%%%%%%%%%%%%%%%%%%%%%%%%%%%%%%%%%%%%%%%%%%%%%%%%%%%%%%%%%%
\begin{figure}
\plotone{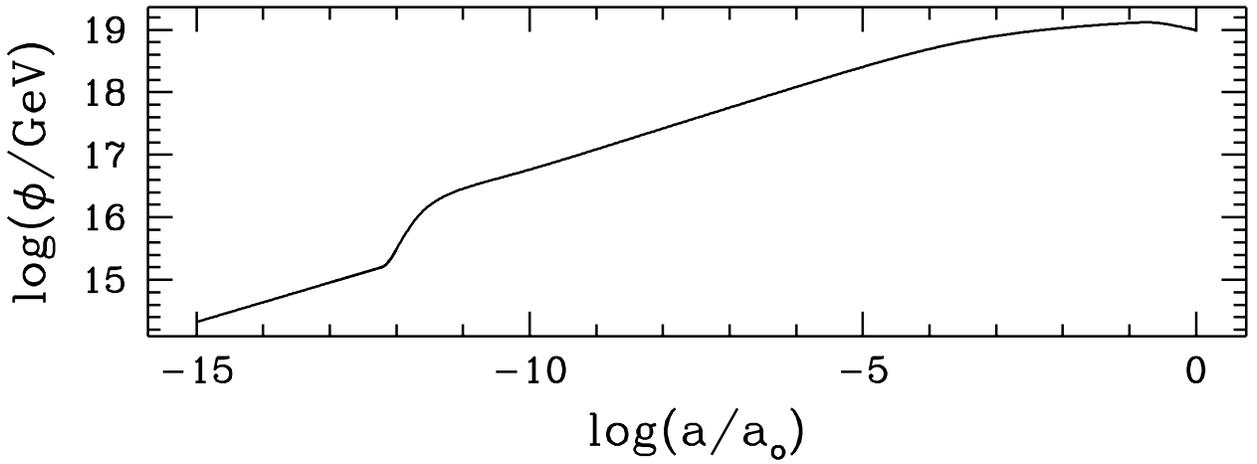}
\epsscale{1.0}
\caption{Evolution of the DE--field. Notice the jump around the
QH scale and the low--$z$ rebounce.}
\label{phi}
\end{figure}
%%%%%%%%%%%%%%%%%%%%%%%%%%%%%%%%%%%%%%%%%%%%%%%%%%%%%%%%%%%%%%%%%%%%%
%%%%%%%%%%%%%%%%%%%%%%%%%%%%%%%%%%%%%%%%%%%%%%%%%%%%%%%%%%%%%%%%%%%%%
%%%%%%%%%%%%%%%%%%%%%%%%%%%%%%%%%%%%%%%%%%%%%%%%%%%%%%%%%%%%%%%%%%%%%
\subsection{Using the SUGRA potential}
Most above features are general and do not depend on the choice of the
potential $V$. We shall now assume that
\begin{equation}
V(\Phi) = {\Lambda^{\alpha+4} \over \phi^\alpha} 
\exp \left( 4 \pi\phi^2 \over m_p^2 \right)
\label{eq:l1}
\end{equation}
(SUGRA potential; see Brax \& Martin 1999, 2001, Brax, Martin \&
Riazuelo 2000). Apparently $V(\Phi)$ depends on the two parameters
$\Lambda$ and $\alpha$. When they are independently assigned,
$\Omega_{de}$ is also fixed. Here we prefer to use $\Omega_{de}$ and
$\Lambda$ as independent parameters. The latter scale is related to
$\alpha$ as shown in Figure \ref{fal}. This relation is almost
independent from the presence of DE--DM coupling.  The slight shifts
due to the couplings, evaluated through the expressions of this paper,
are just slightly above the numerical noise and are also shown in
Figure \ref{fal}.

The potential (\ref{eq:l1}) does not depend on $\theta$ and,
in the radiation dominated era, admits the tracker solution
\begin{equation}
\phi^{\alpha+2} = g_\alpha \Lambda^{\alpha+4} a^2 \tau^2 ~,
\label{eq:l2}
\end{equation}
with $g_\alpha = \alpha (\alpha+2)^2/4(\alpha+6)$.  This tracker
solution, characterizing SUGRA models at very high $z$, is abandoned,
because of the DE--DM coupling, when the term $\phi \dot \theta^2$
exceeds $a^2 V'$, and the field enters a different tracking regime:
\begin{equation}
\phi^{2} = {1 \over 2} \rho_{dm} a^2 \tau^2 ~.
\label{eq:l3}
\end{equation}
Fig.~\ref{f01} is a landscape behavior of densities, starting from the
high--$z$ tracking regime (\ref{eq:l2}), passing then to the new
intermediate tracking regime, and reaching the regimes when DE density
eventually exceeds first radiation ($z \sim 100$), then baryons (at $z
\sim 10$) and DM (at $z \sim 3$).  In Fig.~\ref{phi}, we show the
evolution of the DE--field $\phi$. In Fig.~\ref{f00}, we focus on the
Q--H transition and magnify the scale dependence of the mass and the
onset of $\theta$ oscillations. Finally, in Fig.~\ref{f02}, we show
the behaviors of the density parameters $\Omega_i$ ($i =
r,~b,~\theta,~\phi$, {\it i.e.}  radiation, baryons, DM, DE) {\it
vs.}~the scale factor $a$. More detailed pictures of the single
transitions are shown in MB.

\subsection{Parameter fixing}
In general, once $\Omega_{o,de}$ and $h$ are given, a model with
dynamical (coupled or uncoupled) DE is not yet determined, as $\alpha$
or $\Lambda$ are still to be fixed. When fitting WMAP data, here
below, we shall mostly refer to the energy scale $\Lambda$ and remind
that, when $\Omega_{o,de}$ and $\Lambda$ are given also $\alpha$ is
fixed.

Other potentials show similar features and this implies that, in
general, dynamical DE models depend on one more parameter than, e.g.,
$\Lambda$CDM; accordingly, one can expect that they more easily
accommodate observational data. In principle, data fitting is even
more facilitated by DM--DE coupling, which adds a further parameter
(the strength of coupling) to the theoretical plot.

In the dual--axion model with a SUGRA potential, no such arbitrariness
exists. If this model, where a suitable DM--DE coupling is present,
succeeds in fitting observational data, this will not be favored by
extra parameters. The coupling, in particular, depends on no parameter
and the very scale $\Lambda$ is set by theoretical consistency
arguments, similarly to what happens for the $F_{PQ}$ scale in the PQ
approach.

Let us follow the behavior of $\rho_{dm}$, backwards in time, until
the very beginning of the oscillatory regime, when the approximation
$\theta \ll 1$ begins to hold. Using the scaling law (\ref{eq:m8})
until then and, at earlier times, the more general laws (\ref{enpre}),
together with eq.~(\ref{eq:o1}), we build a system yielding the
scale factor $a_h$ when the fluctuations start, and the scale
$\Lambda$ in the SUGRA potential (\ref{eq:l1}).

It turns out that $a_h$ and $\Lambda$ are almost independent from
$\Omega_{o,dm}$ and $h$, and we shall now outline the qualitative
reasons of this feature.  According to eqs.~(\ref{enpre}) and
(\ref{eq:o2}), at the $a_h$ scale it must be
\begin{equation}
\rho_{h,dm} \simeq  m^2_h\, \phi^2(a_h)~\epsilon^2(a_h)
= \left( m_h \over m_o \right)^2 \mu_{QCD}^4 \, \epsilon^2(a_h)~,
\end{equation} 
where $m_h = m(a_h)$, while $\epsilon^2 \equiv 2 \langle (1 - \cos
\theta)\rangle$ is $\sim \theta^2$  when the oscillatory regime is onset.
 Owing to eq.~(\ref{eq:o1}), however, the oscillation onset
occurs when the axion mass is approximately the inverse of $a_h
\tau(a_h)$ and the latter quantity can be related to the temperature
$T_h$, via the Friedmann equation. This yields that $ T_h^2 \simeq m_p
m_h/8$ and, using again eq.~(\ref{eq:o2}), that
\begin{equation}
T_h \simeq 5 \cdot 10^{-2} \mu_{QCD} \left( {m_p \over \phi_h}
{m_h \over m_o} \right)^{1/2}~.
\end{equation}
Let us now assume that it is now $\phi \sim m_p$ and use the law
(\ref{eq:m8}) from now to $a_h$. It is then
\begin{equation}
T_h \simeq \mu_{QCD} \left[ {5 \cdot 10^3 \over \Omega_{o,dm}
h^2} \left(m_h \over m_o \right)^3 \epsilon^2(a_h) \right]^{1 \over 5}~,
\label{1o5}
\end{equation}
provided that $\mu_{QCD} \simeq 4 \cdot 10^{11} T_o$
and $\rho_{o,dm} \simeq 3 \cdot 10^4 T_o^4 \Omega_{o,dm} h^2$.

The temperature $T_h$ (and the scale $a_h$) are then essentially model
independent because of the power 1/5 at the l.h.s. of eq.~(\ref{1o5}).
If we consider the plots in Fig.~\ref{f00}, obtained through a
numerical integration for $\Omega_{o,dm} = 0.27$ and $h=0.7$, we see
that $(m_h/m_o)^3 \epsilon^2 \sim 10^{-2}$--$ 10^{-3}$, so that $T_h$
falls around $\Lambda_{QCD}$.  However, no appreciable displacement
can be expected just varying $\Omega_{o,dm} $ and $h$, even though we
suppose that this induces substantial variations of the factor
$\left(m_h / m_o \right)^3 \epsilon^2(a_h)$ which, however, cannot
exceed unity or lay below $\sim 10^{-3}$.

The model independence of $a_h$ implies that the scale $\Lambda$ is
also almost model independent.  The numerical result $a_h \simeq
10^{-13}$ is consistent with $\Lambda \simeq 1.5 \cdot
10^{10}$GeV, but neither value will change much just by varying
$\Omega_{o,dm}$. In practice, when $\Omega_{o,dm}$ goes from 0.2 to 0.4,
$\log_{10}(\Lambda/ {\rm GeV})$ (almost) linearly runs from 10.05 to
10.39$\, $ and $a_h$ steadily lays at the eve of the quark--hadron
transition.

\begin{figure}
\plotone{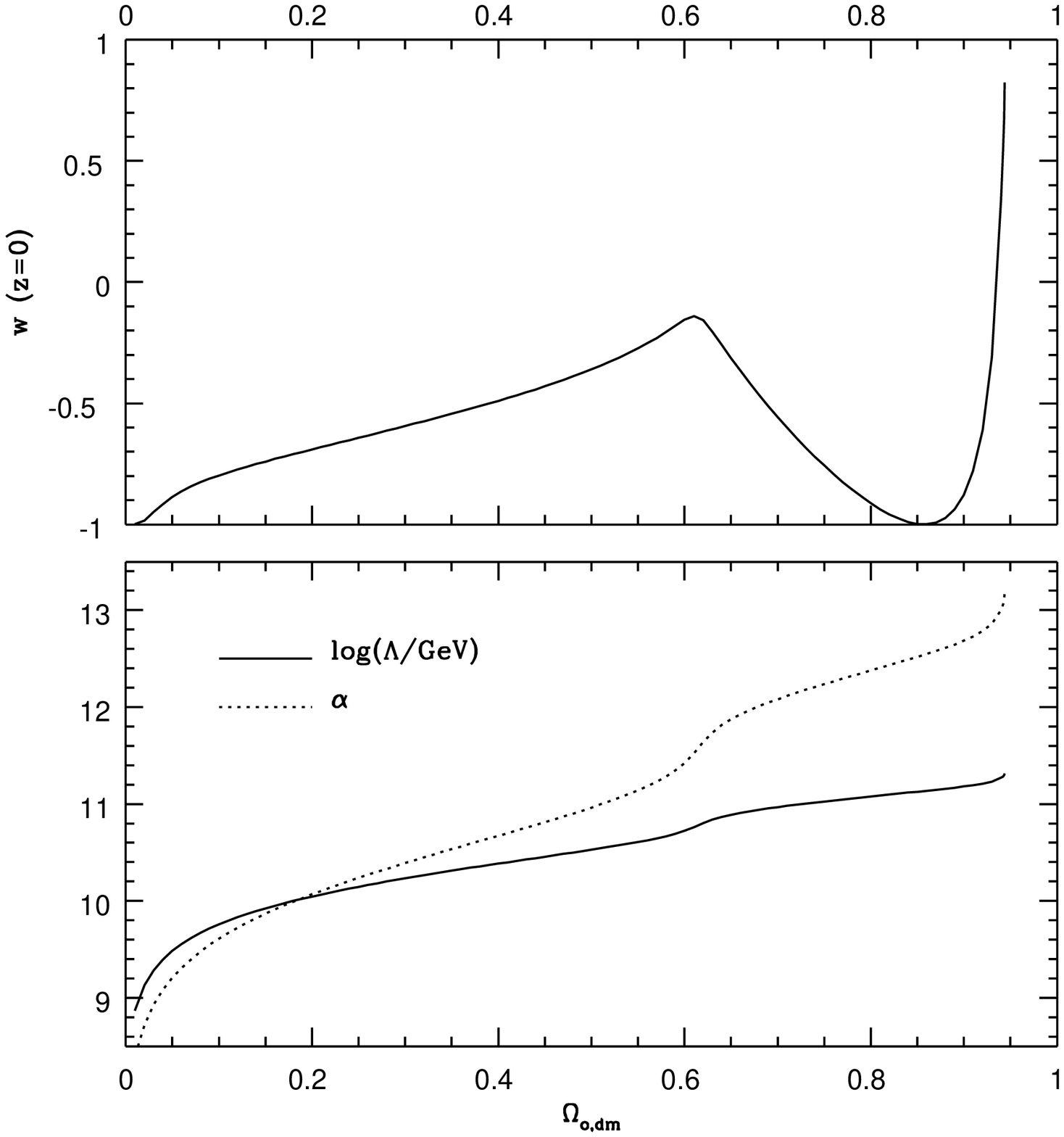}
\epsscale{1.0}
\caption{The lower plot shows the variation of the exponent $\alpha$
and the $\Lambda$ scale in the SUGRA potential, when varying the
present fraction of DM.  The faster $\Lambda$ increase, at
$\Omega_{o,dm} \simeq 0.6$, is due to a stationary behavior in the set
of models where the $\phi$ fields attains (approximately) today the
minimum of the SUGRA potential. In fact, if it were $\Omega_{o,dm}
\simeq 0.6$ today, the upper plot shows that $w$ would corresponds to
top kinetic energy and minimum potential energy. These plots are
obtained for $h = 0.7$. }
\label{lambda}
\vskip -2.truecm
\end{figure}

The residual dependence of $\alpha$ and $\Lambda$ on $\Omega_{o,dm}$
is plotted in Fig.~\ref{lambda}, together with the corresponding
values for $w$ at $z=0$. Significantly smaller values for $\Lambda$
are obtainable only for unphysically small DM densities. 

The only way to modify the result is to consider values of $\phi$
significantly different from $m_p$, today. In fact, the rest of the
plot can be understood as an effect of the position reached by the
$\phi$ field, in the SUGRA potential, at the present time. The peak of
$w$ corresponds to a maximum of kinetic energy and occurs if the
minimum of SUGRA potential is attained today. Today lays in the
proximity of the minimum for a fairly wide $\Lambda$
interval. Accordingly, a significant $\Lambda$ variation, thereabout,
yields just a modest shift of $\Omega_{o,dm}$, as is shown in the
lower plot. For still greater $\Lambda$'s, $\phi$ would still be in
its pre--minimum descent.  For even smaller amounts of DE, the present
$\phi$ configuration would still lay in the $\phi$-MD era, when
kinetic energy dominates for DE, although the very DE contribution to
the overall energy density becomes negligible.
%%%%%%%%%%%%%%%%%%%%%%%%%%%%%%%%%%%%%%%%%%%%%%%%%%%%%%%%%%%%%%%%%%%%%
%%%%%%%%%%%%%%%%%%%%%%%%%%%%%%%%%%%%%%%%%%%%%%%%%%%%%%%%%%%%%%%%%%%%%

A model with DE and DM given by a single complex field, based on a
SUGRA potential, therefore bears a precise prediction on the scale
$\Lambda$, for the observational $\Omega_{o,dm}$ range. Moreover, just
the observational $\Omega_{o,dm}$ range (0.2--0.4) corresponds to
$\phi \sim m_p$ today.

The rest of this paper is devoted to a comparison of this and other
models against WMAP data. In such comparison, however, the scale
$\Lambda$ shall not be fixed, {\it a priori}, in the expected range.
In general, the dual axion model belongs to a wider class of coupled
DM--DE models, which can be dubbed $\phi^{-1}$--models because of the
shape of the coupling, where the scale $\Lambda$ is left free.

The features of $\phi^{-1}$ models, in general, are quite reasonable.
They have two independent dark components, DM and DE, suitably
coupled, about whose nature no assumption is made, just as in standard
coupled DE models. In the latter models, however, the strength of the
coupling is gauged by an extra parameter $\beta$.  $\phi^{-1}$ models,
from this point of view, already offer less freedom, as $C \equiv
\phi^{-1}$ and no coupling modulation is allowed.

Should the comparison of $\phi^{-1}$ models with WMAP data favor
parameter values compatible with cosmological observables and values
of $\Lambda$ compatible with the dual--axion model, this means that
WMAP data support the origin of DE and DM that the dual--axion model
suggests.

\section{Comparison with WMAP data}
WMAP data have been extensively used to provide tight constraints on
cosmological parameters. They consists of high precision estimates of
the anisotropy power spectrum $C_l^T$ up to $l \sim 900$, as well of
the TE correlation power spectrum $C_l^{TE}$ up to $l \sim 450$. We
shall use these data to constrain possible cosmologies, in a parameter
space of 7 to 8 dimensions. A grid-based likelihood analysis would
then require prohibitive amounts of CPU time and we use a Markov Chain
Monte Carlo (MCMC) approach, as it has become customary for CMB
analysis \citep[e.g.,][]{christensen, knox, lewis, kosowski, dunkley}.

The principal analysis of WMAP first--year data \citep{Spergel}
constrained flat $\Lambda$CDM models defined by six parameters:
$\Omega_{o,b} h^2$, $\Omega_{o,m} h^2$, $h$, $n$, the fluctuation
amplitude $A$ and the optical depth $\tau$. Notice that, with the
naming convention used here, $\Omega_{o,m} \equiv \Omega_{o,b} +
\Omega_{o,dm}$. As possible extensions of $\Lambda$CDM cosmologies,
several works considered models with a fixed state parameter $w \equiv
p_{de} / \rho_{de}$ \citep[e.g.,][]{Spergel,bean,teg:04,melchiorri},
or adopted $z$--dependent parameterizations of $w(z)$ interpolating
between early--time and late--time values
\citep[e.g.,][]{corasaniti,jassal,rapetti04}. A general conclusion was
that current data mostly allow to constrain only the present state
parameter, $w(z=0) \lesssim -0.80$.

In this work we consider, instead, three classes of dynamical DE,
requiring the introduction of additional parameters specifying the
physical properties of the scalar field. (i) SUGRA dynamical DE
require the introduction of $\lambda = \log_{10}(\Lambda/{\rm GeV})$,
yielding the energy scale in the potential (12). (ii) In constant
coupling DE, the coupling parameter $\beta = C\,(3 m_p^2 /16
\pi)^{1/2}$ is also needed. (iii) In the case of the dual--axion
model, the last parameter is excluded, for it is simply $C =
\phi^{-1}$.  Also the scale $\Lambda$ is constrained by the
requirement that $\Omega_{o,de}$ lays in a fair range (also solving 
the strong $CP$ problem). Hence, in the dual--axion model, $\Lambda$
and $\Omega_{o,de}$ are no longer independent parameters. We however
consider a wider class of coupled DE models, that we call $\phi^{-1}$
models, leaving $\lambda$ as a free parameter. Our aim is to test
whether WMAP data constrain it into the region turning a $\phi^{-1}$
model into a dual--axion model.

In the use of MCMC, as well as in any attempt to fit CMB data to
models, a linear code providing $C_l$'s is needed. Here we use our
optimized extension of CMBFAST \citep{cmbfast}, able to inspect the
cosmologies (i), (ii) and (iii). Then, the likelihood of each model is
evaluated through the publicly available code by the WMAP team
\citep{wmap:verde} and accompanying data
\citep{wmap:hinshaw,wmap:kogut}.

\subsection{Implementing a MCMC algorithm}
A MCMC algorithm samples a known distribution ${\cal L}({\bf x})$ by
means of an arbitrary trial distribution $p({\bf x})$. Here $\cal L$
is a likelihood and ${\bf x}$ is a point in the parameter space. The
chain is started from a random position ${\bf x}$ and moves to a new
position ${\bf x}^\prime$, according to the trial distribution. The
probability of accepting the new point is given by ${\cal L}({\bf
x}^\prime)/{\cal L}({\bf x})$; if the new point is accepted, it is
added to the chain and used as the starting position for a new
step. If ${\bf x}^\prime$ is rejected, a replica of ${\bf x}$ is added
to the chain and a new ${\bf x}^\prime$ is tested.

In the limit of infinitely long chains, the distribution of points
sampled by a MCMC describes the underlying statistical process. Real
chains, however, are finite and convergence criteria are critical.
Moreover, a chain must be required to fully explore the high
probability region in the parameter space. Statistical properties
estimated using a chain which has yet to achieve good convergence or
mixing may be misleading. Several methods exist to diagnose mixing and
convergence, involving either single long chains or multiple chains
starting from well separated points in the parameter space, as the one
used here. Once a chain passes convergence tests, it is an accurate
representation of the underlying distribution.

In order to ensure mixing, we run six chains of $\sim 30000$ points
each, for each model category. We diagnose convergence by requiring
that, for each parameter, the variances both of the single chains and
of the whole set of chains ($W$ and $B$, respectively) satisfy the
Gelman \& Rubin test~\cite{wmap:verde,gelrub}, $R < 1.1$ with:
\begin{equation}
R = {[(N-1)/N]W +(1 +1/M)B \over W}~. 
\end{equation}
Here each chain has $2N$ points, but only the last $N$ points are used
to estimate variances, and $M$ is the total number of chains. In most
model categories considered, we find that the slowest parameter to
converge is $\lambda$.
 
\subsection{Results}
The results of this paper mostly concern the fit of the double--axion
model with WMAP data. Let us however also briefly outline the features
of the model.

Quite in general, cosmological models fitting observations require a a
triple coincidence between DM, baryon and DE densities, the last one
occurring just at the present epoch. Any such coincidence requires the
tuning of a suitable parameter. It is then natural to look for an
underlying physics, able to predict the parameter values that
cosmology requires.

If DM is made of axions, we must tune the $F_{PQ}$ parameter in the NG
potential involving a $\Phi$ field, whose phase $\theta$ is then the
axion field. Here we saw that the PQ model does not strictly require a
NG potential. A potential causing no immediate settling at an energy
minimum, but inducing a slow rolling along a tracking solution,
achieves the same aims. If such potential is SUGRA, it contains an
energy scale $\Lambda$ to be tuned, in stead of $F_{PQ}$. If this is
done, so to yield the required amount of DM, this scheme provides, as
an extra bonus, a fair amount of DE, which is prescribed to interact
with DM in a peculiar way, never studied in the literature.

The possible shapes of DE--DM interaction considered up to now
involved an {\it ad--hoc} coupling parameter. No such parameter exists
here. For instance, a standard model of coupled SUGRA involves a
density parameter $\Omega_{o,dm}$, an energy scale $\Lambda$ and a
coupling parameter $\beta$. Here, not only $\beta$ no longer exists,
but also $\Lambda$ is fixed once $\Omega_{0,dm}$ is assigned.  The
model is therefore highly constrained and data can easily falsify it.
The only available degree of freedom, to modify it, amounts
to replacing the SUGRA potential with another potential shape.
Another reserve to be born in mind is that axion models have a
contribution to DM coming from the decay of topological structures,
which were not considered here.

\begin{figure}
\plotone{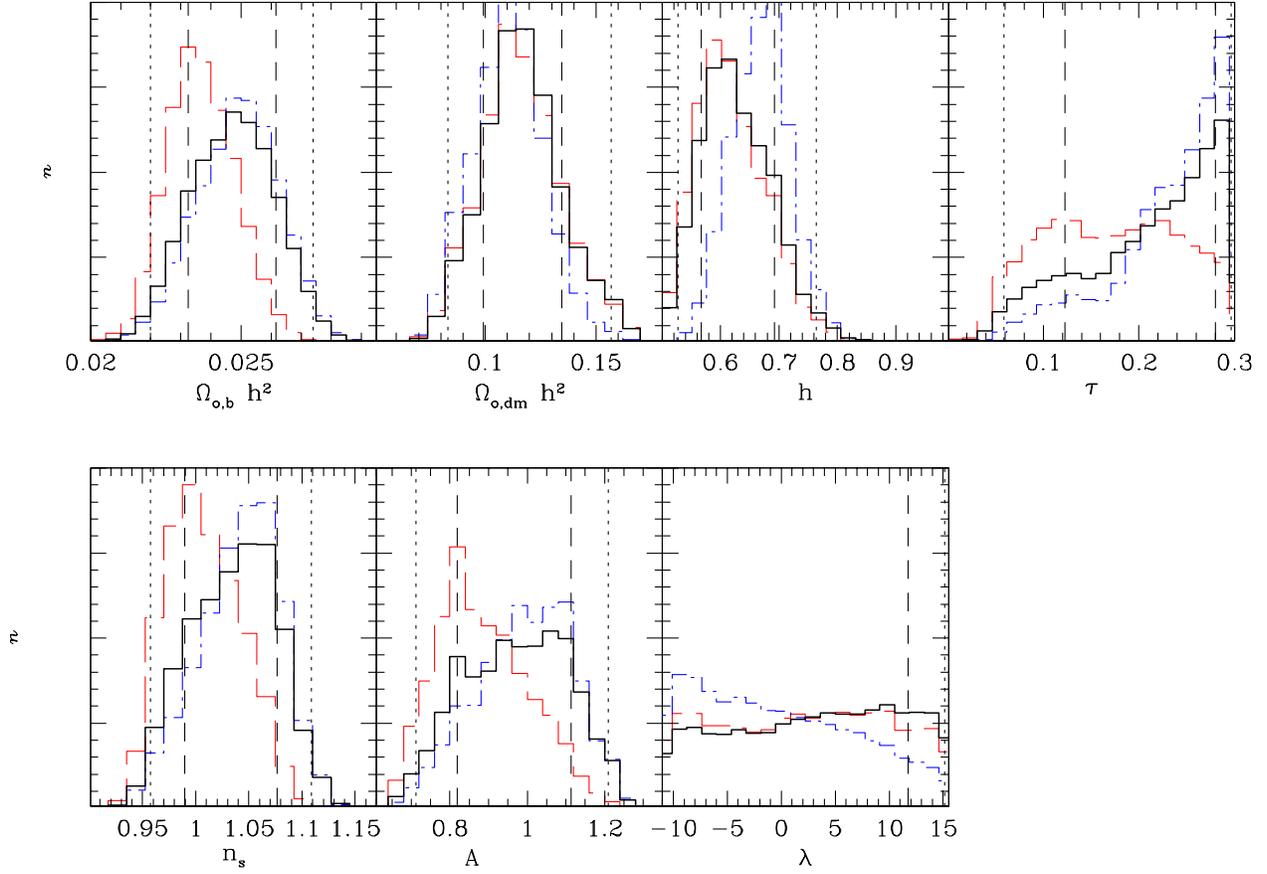}
\epsscale{1.0}
\caption{Marginalized distributions for the 7--parameters SUGRA model
with no priors (solid lines), BBNS prior (long dashed) or HST prior
(dot--dashed).  Short dashed (dotted) vertical lines show the
boundaries of 68.3 \% c.l.  (95.4 \% c.l.) interval; for $\lambda$
only upper limits are shown.}
\label{fig:pdf1D1}
\end{figure}

\begin{figure}
\plotone{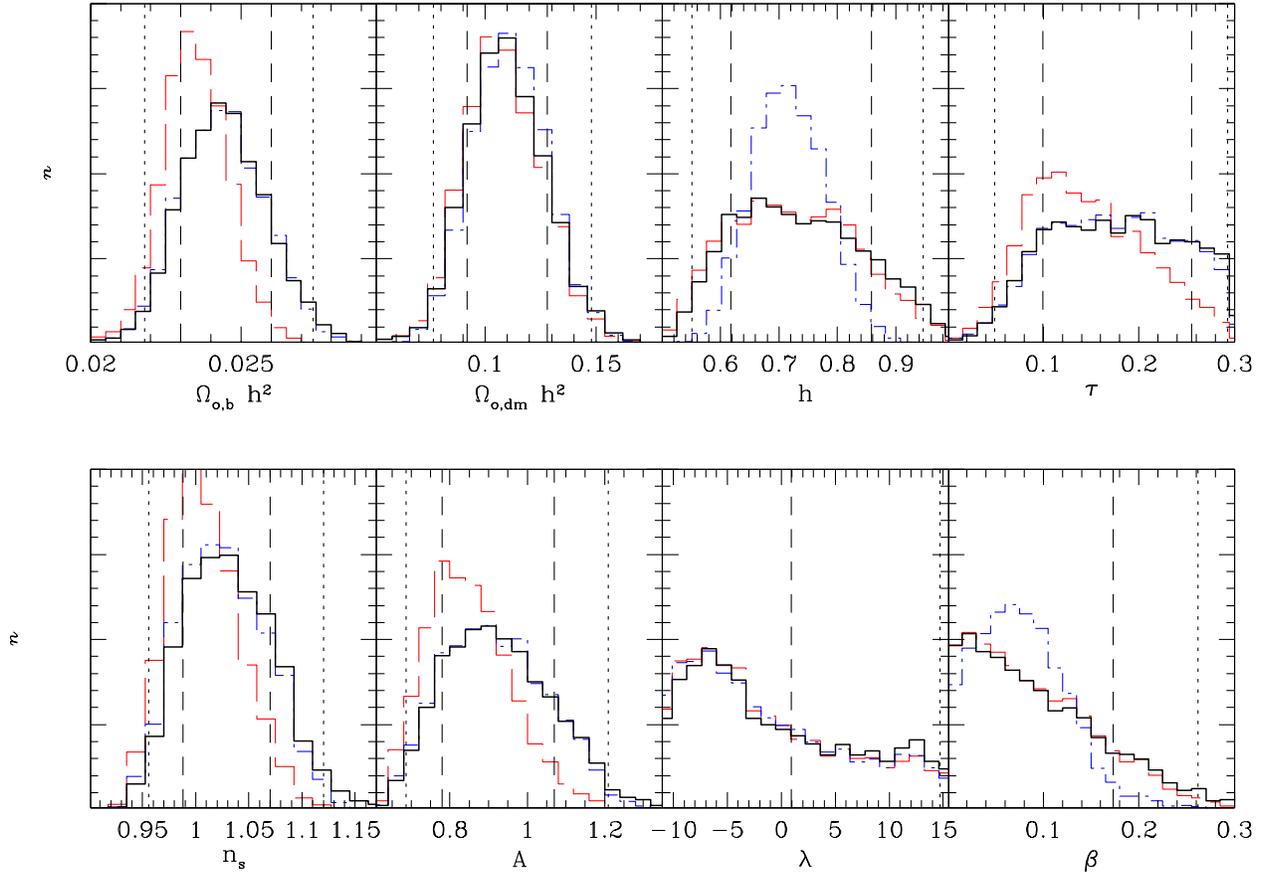}
\epsscale{1.0}
\caption{As Fig.~\ref{fig:pdf1D1} but for the 8--parameters constant
coupling model. For $\lambda$ and $\beta$ only the upper
c.l. boundaries are shown.}
\label{fig:pdf1D2}
\end{figure}

\begin{figure}
\plotone{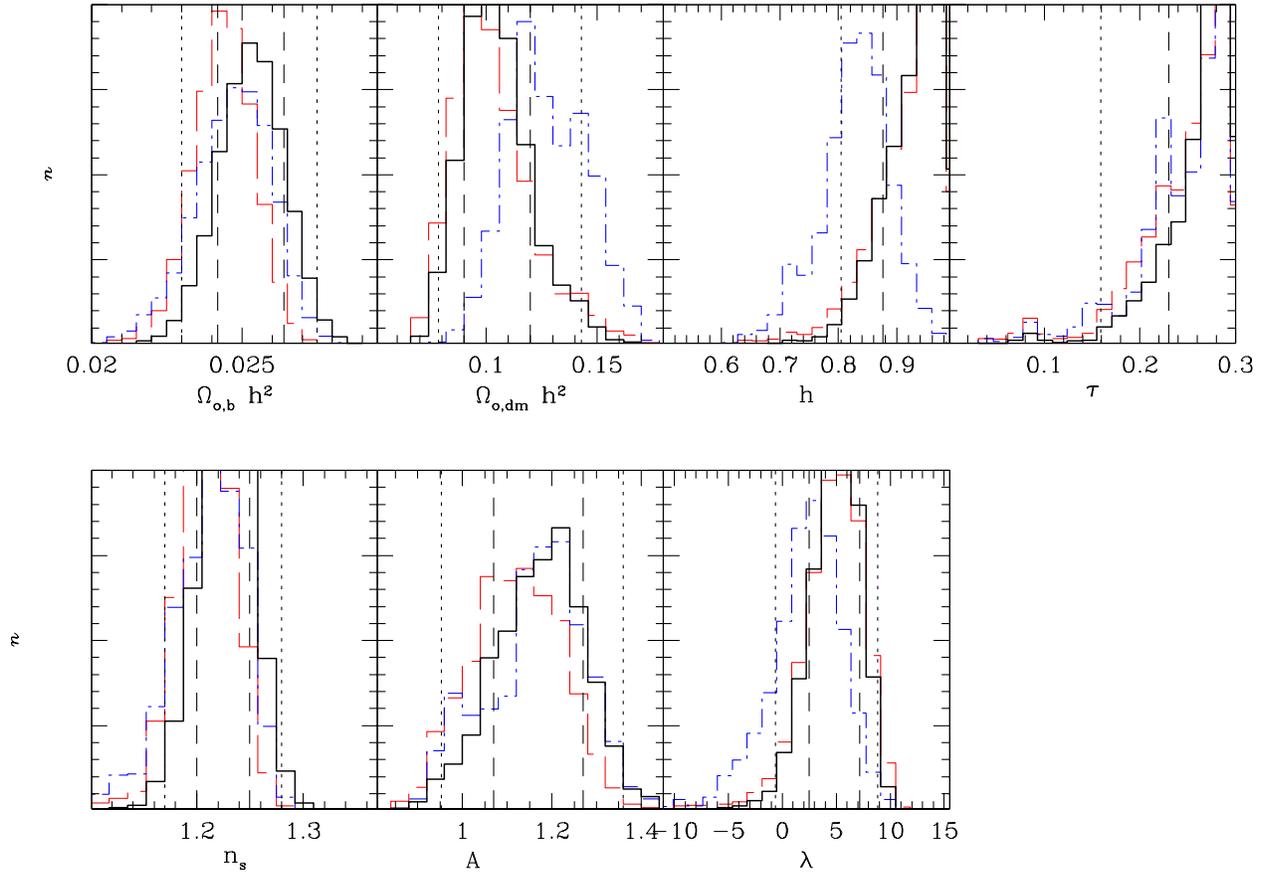}
\epsscale{1.0}
\caption{As Fig.~\ref{fig:pdf1D1} but for the 7--parameters $\phi^{-1}$
model.}
\label{fig:pdf1D3}
\end{figure}

\begin{figure}
\plotone{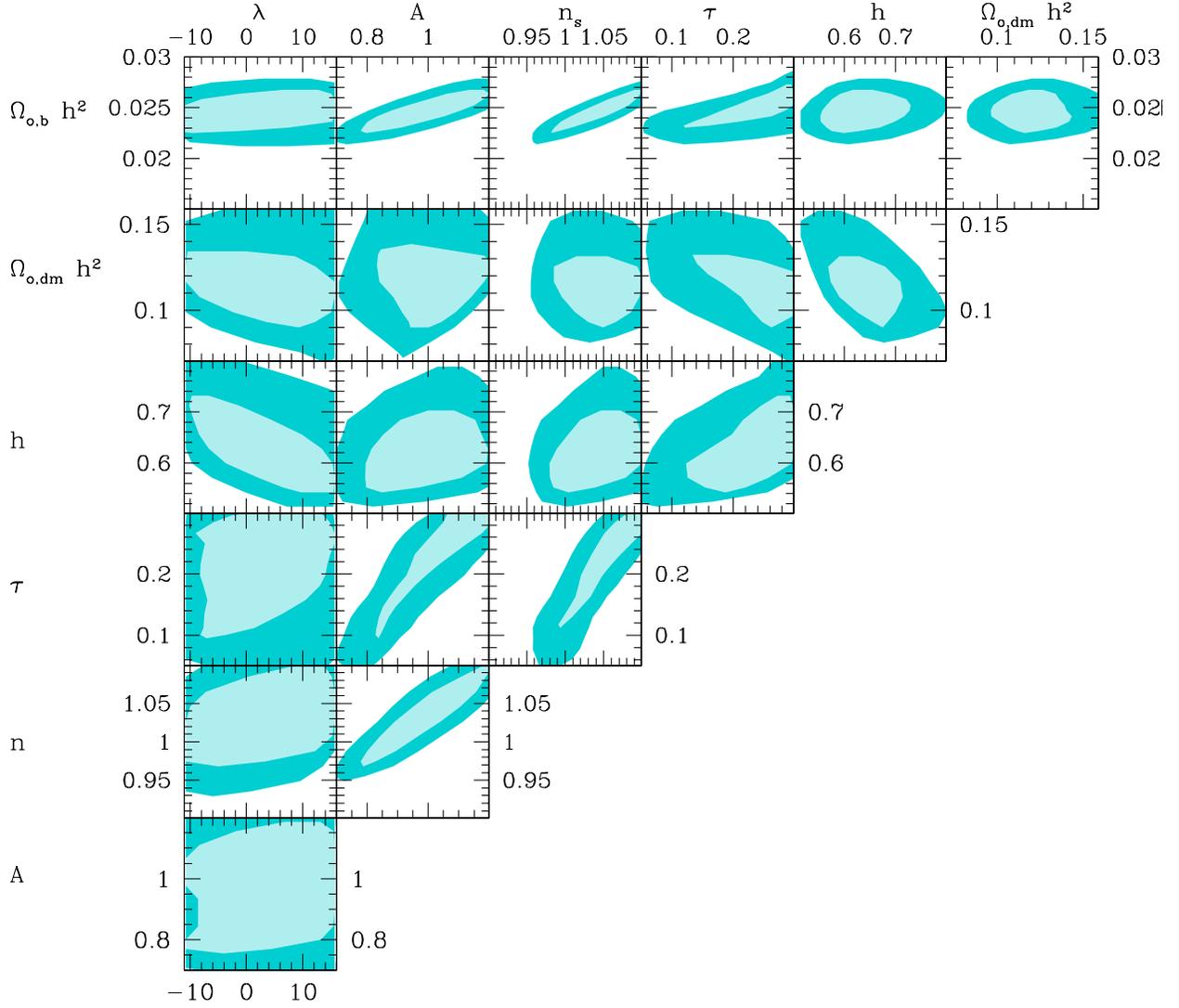}
\epsscale{1.0}
\caption{Joint 2D constraints for SUGRA models. 
Light (dark) shaded areas delimit the region enclosing 68.3 \%
(95.4 \%) of the total points.}
\label{fig:pdf2D1}
\end{figure}

\begin{figure}
\plotone{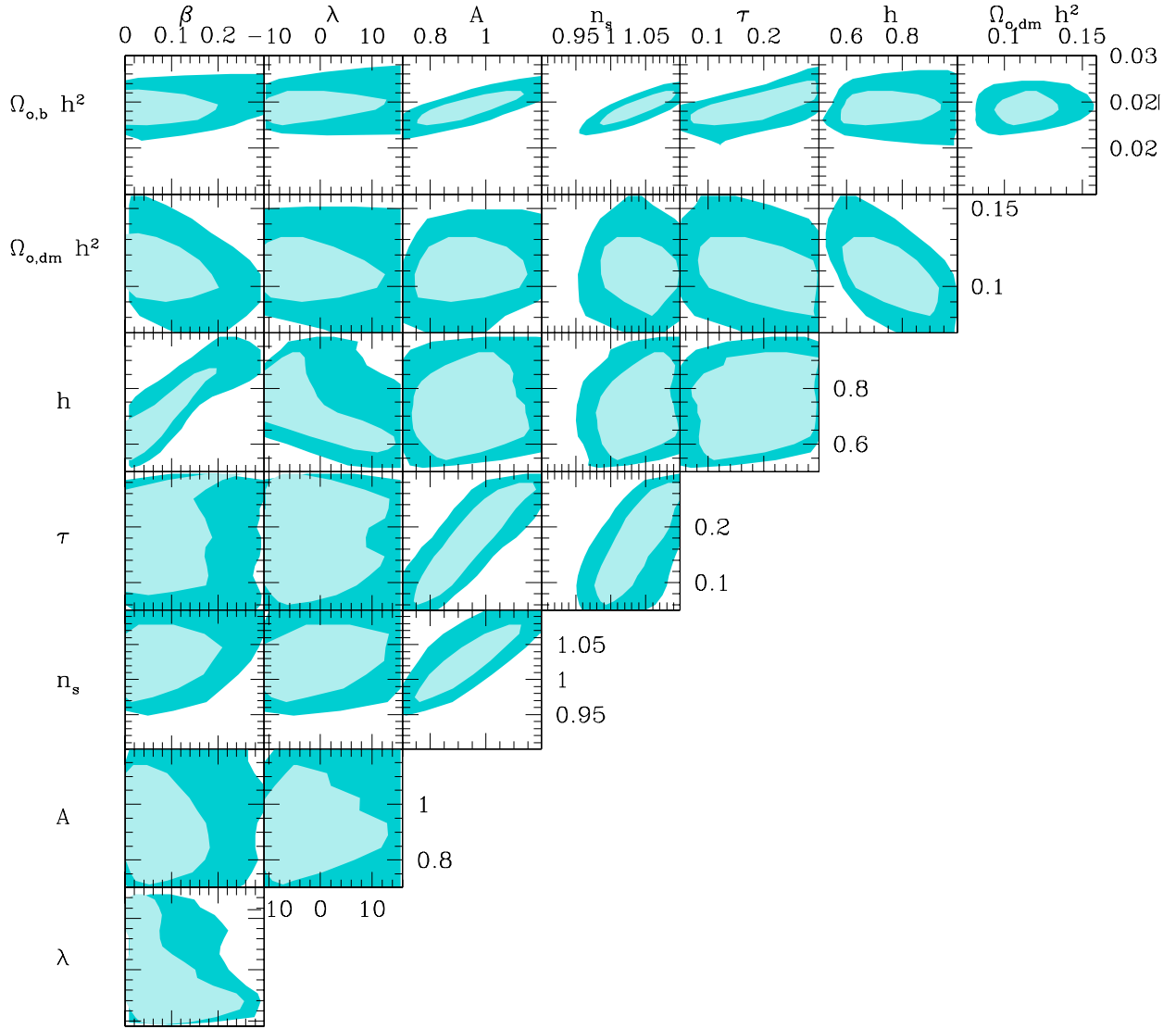}
\epsscale{1.0}
\caption{As Fig.~\ref{fig:pdf2D1} but for constant coupling models.}
\label{fig:pdf2D2}
\end{figure}

\begin{figure}
\plotone{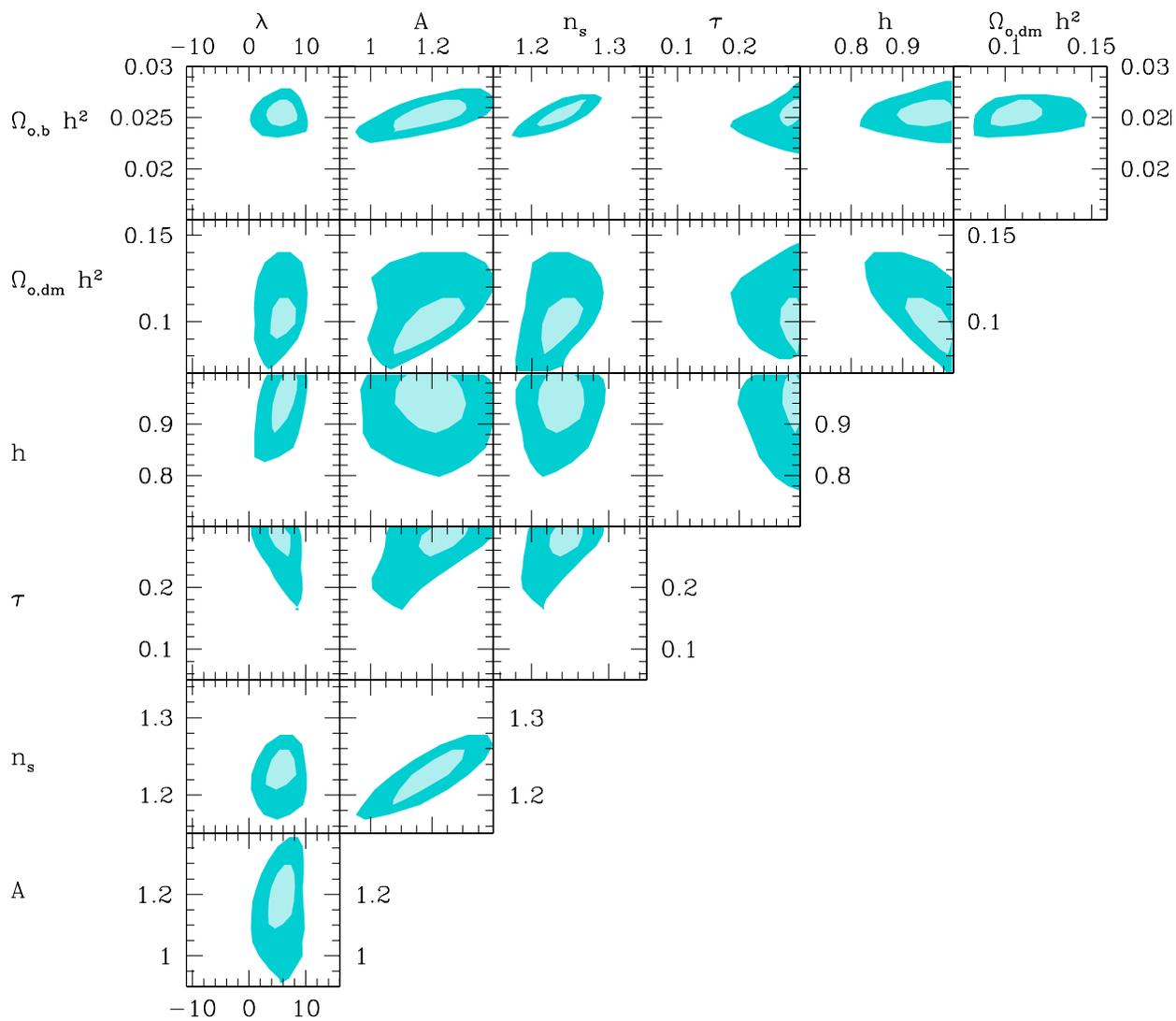}
\epsscale{1.0}
\caption{As Fig.~\ref{fig:pdf2D1} but for $\phi^{-1}$ models. Here,
cosmological parameters are more stringently constrained than in other
models.}
\label{fig:pdf2D3}
\end{figure}

These arguments outline the significance of the fit of the model with
WMAP data. The basic results of this work concern this fit and are
summarized in the Tables~\ref{tab:res1}--\ref{tab:res3}. For each
model category we list the expectation values of each parameter and
the associated variance; we also list the values of the parameters of
the best fitting models.  The corresponding marginalized distributions
are plotted in figures~\ref{fig:pdf1D1}--\ref{fig:pdf1D3}, while joint
2D confidence regions are shown in
figures~\ref{fig:pdf2D1}--\ref{fig:pdf2D3}.

The values of $\chi^2$, for each category of models can be compared,
taking into account the number of degrees of freedom.  This comparison
is shown in Table~\ref{tab:chi2}. The smallest $\chi^2$ is obtained
for the uncoupled SUGRA model, which performs slightly better than
$\Lambda$CDM. Differences however are really small and yield no
support to any model category.

It must be however reminded that the $\phi^{-1}$ models, whose fitting
results are reported in Table~3 and Figures \ref{fig:pdf1D3} and
\ref{fig:pdf2D3}, include the dual--axion model, but many other cases
as well. Our approach was meant to test whether CMB data carry
information on $\lambda$ and how this information fits the $\lambda$
range turning a $\phi^{-1}$ model into the dual--axion model.

Let us also outline that Tables 1 and 2 and the corresponding figures,
{\it concerning uncoupled or constant--coupling} SUGRA models, outline
that WMAP data provide no real constraint on $\lambda=\log(\Lambda/
{\rm GeV})$, when allowed to vary from $\sim -12$ to 16. No limitation
exists even on its sign. On the contrary, when a $\phi^{-1}$ coupling
is set, loose but precise limitations on $\lambda$ arise, as is shown
in Table 3. In the presence of this coupling, the 2--$\sigma$
$\Lambda$--interval ranges from $\sim 10$ to $\sim 3 \cdot
10^{10}$GeV, {\it including the range required by the dual--axion
model}.
%---------------------------------------------------------------------------
\begin{table}
\caption{SUGRA parameters in the absence of DE--DM coupling: 
for each parameter $x$, the expectation
 value $\langle x \rangle$, variance $\sigma_x$, and maximum
 likelihood values $x_{max}$, in the 7--dimensional parameter space,
 are shown.} 
\label{tab:res1}
\vglue 0.2truecm
\begin{center}
\begin{tabular}{cccc}
\hline
\rule[-1ex]{0pt}{3.5ex}
      $ x $    &  $ \langle x \rangle$  &  $\sigma_x$ &  $x_{max}$  
  \\
\hline
$\Omega_{o,b} h^2$  &     0.025  &   0.001  &    0.026  \\
$\Omega_{o,dm} h^2$  &     0.12   &   0.02   &    0.11   \\
$ h $           &     0.63   &   0.06   &    0.58   \\
$ \tau$         &     0.21   &   0.07   &    0.28   \\
$ n_s$          &     1.04   &   0.04   &    1.08   \\
$ A $           &     0.97   &   0.13   &    1.11   \\
$\lambda$       &     3.0    &    7.7   &    13.7   \\
\hline
\end{tabular}
\end{center}
\vglue -0.5truecm
\caption{SUGRA parameters in the presence of a constant DE--DM
coupling $\beta$: The parameter space is 7--dimensional and parameter
values are shown as in the previous Table.} 
\label{tab:res2}
\vglue0.2truecm
\begin{center}
\begin{tabular}{cccc}
\hline
\rule[-1ex]{0pt}{3.5ex}
      $ x $    &  $ \langle x \rangle$  &  $\sigma_x$ &  $x_{max}$  
  \\
\hline
$\Omega_{o,b} h^2$  &     0.024  &   0.001  &    0.024  \\
$\Omega_{o,dm} h^2$  &     0.11   &   0.02   &    0.12   \\
$ h $           &     0.74   &   0.11   &    0.57   \\
$ \tau$         &     0.18   &   0.07   &    0.17   \\
$ n_s$          &     1.03   &   0.04   &    1.02   \\
$ A $           &     0.92   &   0.14   &    0.93   \\
$\lambda$       &    -0.5    &    7.6   &    8.3    \\
$\beta$         &     0.10   &   0.07   &    0.07   \\
\hline
\end{tabular}
\end{center}
\vglue -0.5truecm
\caption{SUGRA parameters for a $\phi^{-1}$ model.
The parameter $\lambda$ is left arbitrary; at variance from
other model categories, $\lambda$ here is constrained and
consistency with the dual--axion model is recovered at $2\, \sigma$'s.
Parameter values are shown as in Table 1.}
\label{tab:res3}
\vglue0.2truecm
\begin{center}
\begin{tabular}{cccc}
\hline
\rule[-1ex]{0pt}{3.5ex}
      $ x $    &  $ \langle x \rangle$  &  $\sigma_x$ &  $x_{max}$  
  \\
\hline
$\Omega_{o,b} h^2$  &     0.025  &   0.001  &    0.026  \\
$\Omega_{o,dm} h^2$  &     0.11   &   0.02   &    0.09   \\
$ h $           &     0.93   &   0.05   &    0.98   \\
$ \tau$         &     0.26   &   0.04   &    0.29   \\
$ n_s$          &     1.23   &   0.04   &    1.23   \\
$ A $           &     1.17   &   0.10   &    1.20   \\
$\lambda$       &     4.8    &    2.4   &     5.7   \\
\hline
\end{tabular}
\end{center}
\end{table}
%+++++++++++++++++++++++++++++++++++++++++++++++++++++++++++++++++++
%++++++++++++++++++++++++++++++++++++++++++++++++++++++++++++++++++++++++++
\begin{table}
\caption{Goodness of fit. For each class of DE considered, Table lists
the number of degrees of freedom (d.o.f.), the reduced $\chi^2_{eff}$,
and the corresponding probability of the best--fit model. Figures for
$\Lambda$CDM and dual--axion models are also included. $\Lambda$CDM models have
1342 degrees of freedom, uncoupled and $\phi^{-1}$ have 1341, while
fixed--$\beta$ has 1340.}
\label{tab:chi2}
\vglue0.2truecm
\begin{center}
\begin{tabular}{cccc}
\hline
\rule[-1ex]{0pt}{3.5ex}
                      &  $\chi^2_{eff}$  &   prob.      \\
\hline
no coupling               &  1.064    &     5.0 \%   \\
$\beta$--coupling         &  1.066    &     4.7 \%   \\
$\phi^{-1}$--coupling     &  1.074    &     2.9 \%   \\
dual--axion                        &  1.081    &     2.0 \%   \\
$\Lambda$CDM              &  1.066    &     4.7 \%   \\

\hline
\end{tabular}
\end{center}
\end{table}
%---------------------------------------------------------------------------

%---------------------------------------------------------------------------

%---------------------------------------------------------------------------

%---------------------------------------------------------------------------

%---------------------------------------------------------------------------
%---------------------------------------------------------------------------

%---------------------------------------------------------------------------
\begin{figure}
\plotone{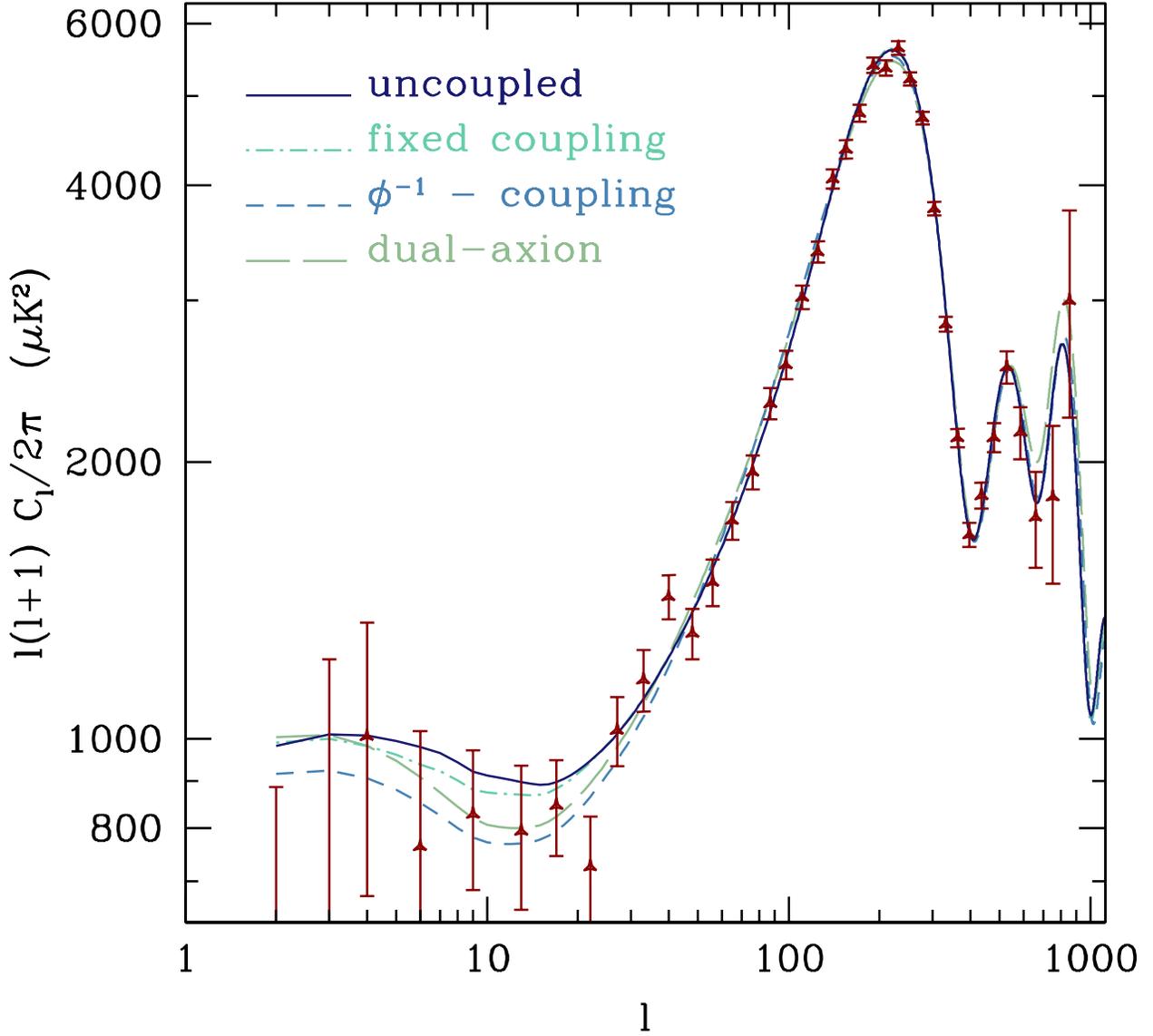}
\epsscale{1.0}
\caption{$C_l^T$ spectra for the best fit SUGRA (solid line), constant
coupling (dotted line), $\phi^{-1}$--coupling (dashed) and dual--axion
(dot--dashed) models. Dual--axion model results from considering only
those $\phi^{-1}$ with $9.5 \le \lambda \le 10.5$. The binned
first--year WMAP data are also plotted.}
\label{fig:taps}
\end{figure}
%---------------------------------------------------------------------------
\begin{figure}
\plotone{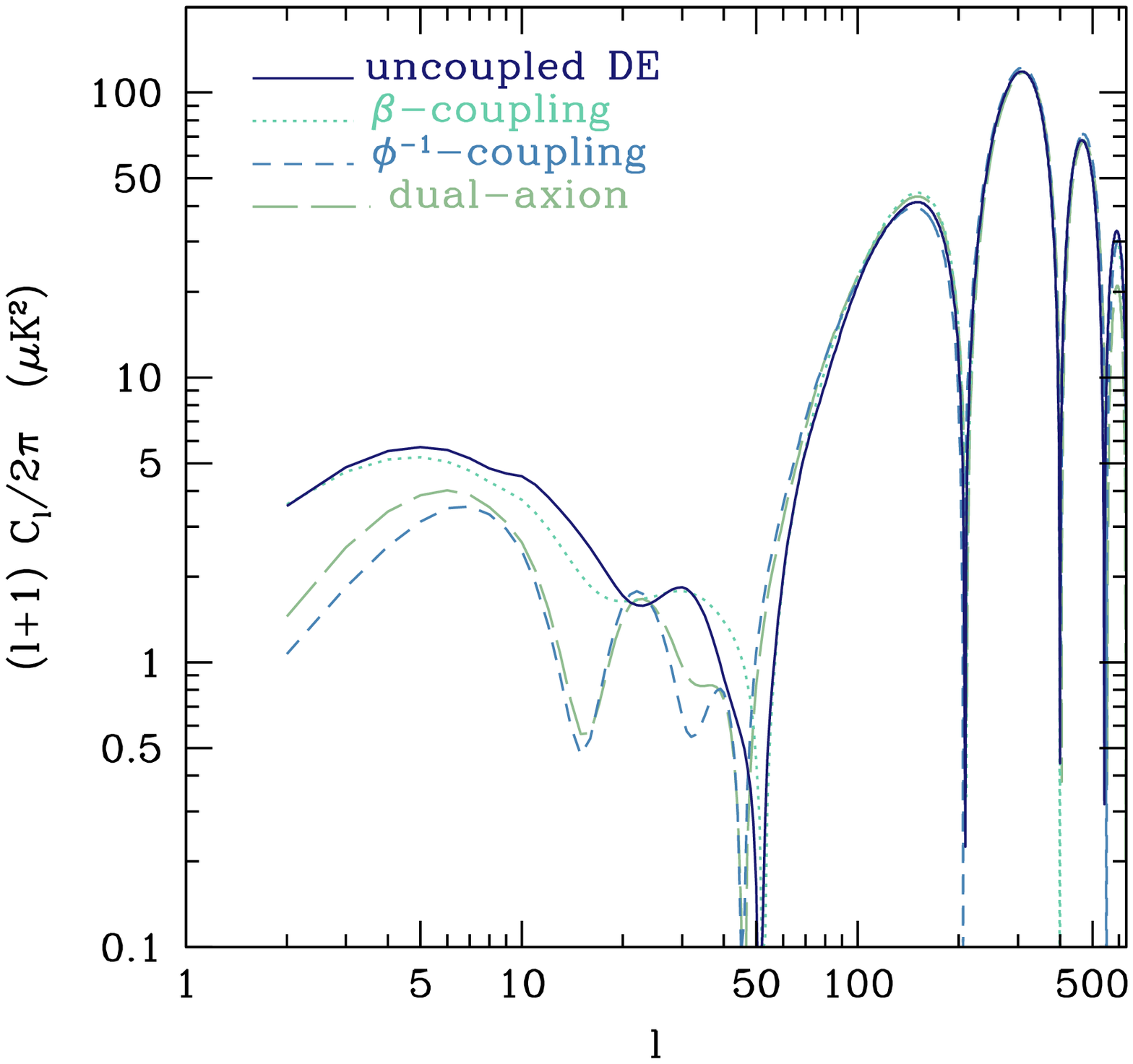}
\epsscale{1.0}
\caption{Best fit $C_l^{TE}$ spectra.}
\label{fig:teaps}
\end{figure}
%---------------------------------------------------------------------------
\begin{figure}
\plotone{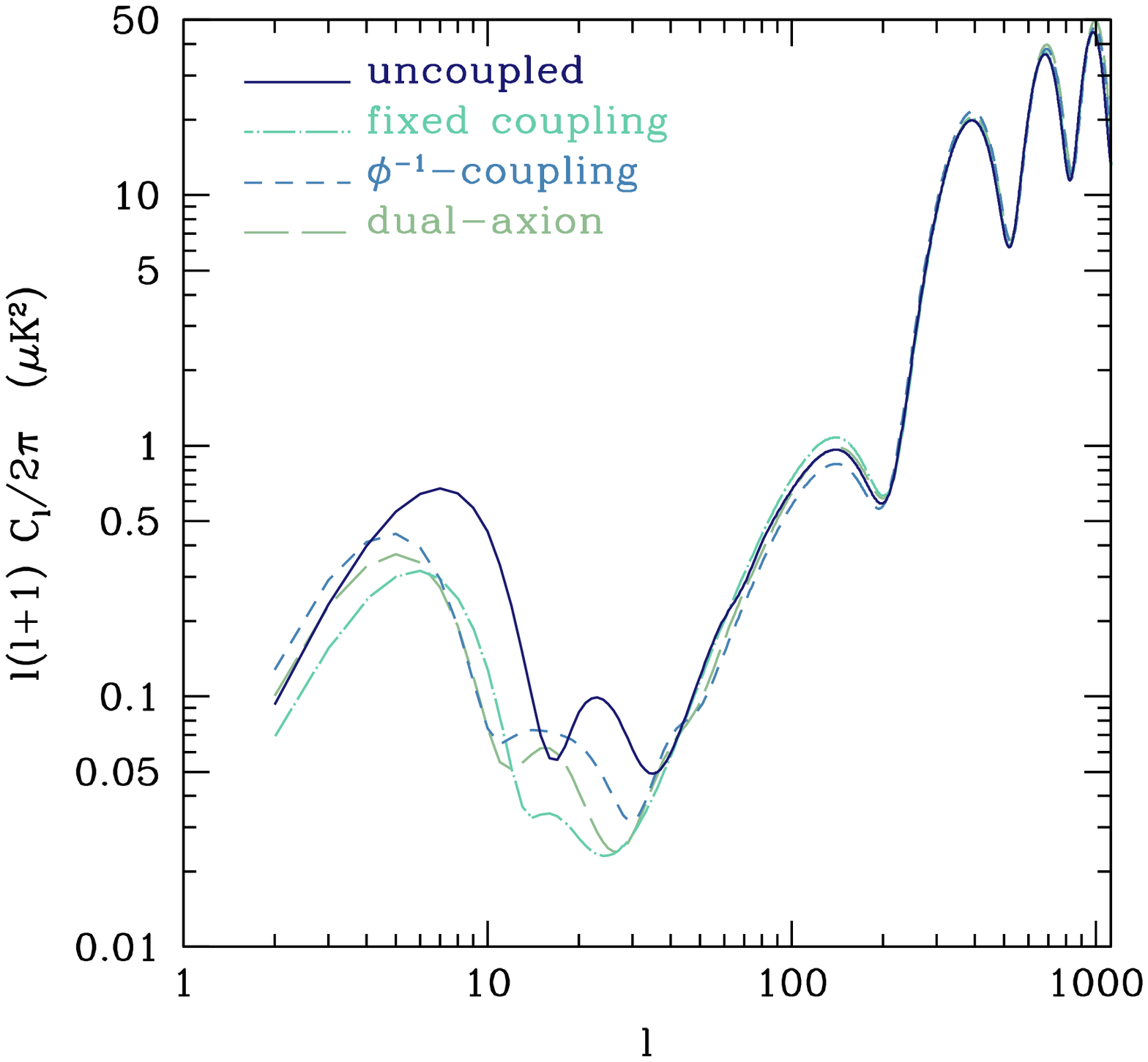}
\epsscale{1.0}
\caption{Best fit $C_l^{E}$ spectra.}
\label{fig:eaps}
\end{figure}
%---------------------------------------------------------------------------

In fact, in this case, the location of peaks in spectra is more
strictly related to the $\phi$ evolution. In turn, only a restricted
sect of $\Lambda$ values allows an $l$--dependence of multipoles
consistent with data. Fig.~\ref{fig:taps} shows this fact for
anisotropy, while Figs.~\ref{fig:teaps} and \ref{fig:eaps} 
are predictions for the TE and E--polarization spectra.

Fig.~\ref{fig:taps} also shows why no model category neatly
prevails. At large $l$ all best--fit models yield similar
behaviors. In turn this shows that discrimination could be achieved
by improving large angular scale observation, especially for
polarization, so to reduce errors on small--$l$ harmonics.

\section{Discussion}
Our analysis does not concern the double axion model only, but the
whole spectrum of models built using a SUGRA potential. This is
useful also to evaluate the significance of the fit of the dual--axion
model.

\subsection{Uncoupled and coupled SUGRA models}
A first point we must therefore outline is that SUGRA {\it uncoupled}
models are consistent with WMAP data. The ratio $w = p/\rho$, for most
these models is $ \lesssim - 0.80$ at $z=0$. However, they exhibit a
fast variation of $w$, which already attains values $\sim -0.6$ at $z
\sim 1$--2. This sharp decrease does not conflict with data and these
models perform even better than $\Lambda$CDM. For uncoupled or
constant--coupling SUGRA models, the analysis in the presence of
priors leads to analogous conclusions.

Best fit cosmological parameters exhibit some dependence on the model.
First, the opacity $\tau$ is pushed to values exceeding the
$\Lambda$CDM estimates \citep[see also][]{corasaniti}.  This can be
understood in two complementary ways: (i) Dynamical DE models, in
general, exhibit a stronger ISW effect, as the field $\phi$ itself
varies during the expansion, and DE effects extend to greater
$z$. This increases $C_l^T$ in the low--$l$ plateau
\cite[e.g.,][]{weller}. To compensate this effect the fit tends to
shift the primeval spectral index $n_s$ to a greater value. Owing to
the $\tau$--$n$ degeneracy, this is then compensated by increasing
$\tau$. (ii) In dynamical DE models, the expected TE correlation, at
low $l$, is smaller than in $\Lambda$CDM (Colombo et al.~2003).  A
given observed correlation level, therefore, requires a greater
$\tau$. In any case, values of $\tau \sim 0.07$ keep consistent with
data within less than 2--$\sigma$'s.

Greater $\tau$'s push upwards $\Omega_b h^2$ estimates, although best
fit values are consistent with $\Lambda$CDM within 1--$\sigma$. Adding
a prior on $\Omega_b h^2 = 0.0214 \pm 0.0020$ (BBNS estimates, see,
e.g.  Kirkman et al. 2003) lowers $h$, within 1--$\sigma$ from HST
findings.  We therefore also consider the effect of a prior on $h$. In
Figures~\ref{fig:pdf1D1} and~\ref{fig:pdf1D2} the effects of priors
are shown by the dashed red line (prior on $\Omega_b h^2$) and the
dot--dashed blue line (prior on $h$).

The former prior affects mainly reionization and $n_s$; $\tau$ and
$n_s$ are lowered to match WMAP's findings, the high tail of $\tau$
distribution is partly suppressed.  The physical analysis of primeval
objects causing reionization~\citep[e.g.,][]{ciardi03,ricotti04}
could however hardly account for values of $\tau \sim 0.3$ which are still
allowed but certainly not {\it required}.

The latter prior favors greater $h$ values. In the absence of
coupling, this favors low--$\lambda$ models, closer to
$\Lambda$CDM. In fact, the sound horizon at decoupling is unaffected
by the scale $\Lambda$, while the comoving distance to last scattering
band is smaller for greater $\lambda$'s. Then, as $\lambda$ increases,
lower $h$ values are favored to match the angular position of the
first peak.  In the presence of coupling, there is a simultaneous
effect on $\beta$, as greater $\beta$'s yield a smaller sound horizon
at recombination, so that the distribution on $h$ is smoother.

A previous analysis of WMAP limits on constant coupling models had
been carried on by Amendola \& Quercellini (2003). Their analysis
concerned potentials $V$ fulfilling the relation $dV/d\phi = BV^N$,
with suitable $B$ and $N$. Furthermore, they assume that $\tau \equiv
0.17$.  Our analysis deals with a different potential and allows more
general parameter variations. The constraints on $\beta$ we find are
less severe. It must be however outlined that $\beta \simgreat
0.1$--0.2 seem however forbidden by a non--linear analysis of
structure formation (Macci\`o et al. 2004).

\subsection{Dual--axion models}
Let us consider then $\phi^{-1}$--models, which generalize dual--axion
models to an arbitrary $\Lambda$ scale, used as a parameter to fit
data. 

Parameters are better constrained in this case, although the overall
model likelihood is similar. This is made evident by
Fig.~\ref{fig:pdf2D3}. In particular, at variance from the former
case, the energy scale $\Lambda$ is significantly constrained and,
within 2--$\sigma$, {\it constraints are consistent with the
double--axion model.} 

Several other parameters are constrained, similarly to coupled or
uncoupled SUGRA models. What is peculiar of $\phi^{-1}$--models is the
range of favored $h$ values: the best--fit 2--$\sigma$ interval does
not extend much below 0.85$\, $.

This problem is slightly more severe for the dual--axion model. The point
is that this model naturally tends to displace the first $C_l^T$ peak
to greater $l$ (smaller angular scales) as coupling does, in any
case. But, in the absence of a specific coupling parameter, the very
intensity of coupling, in these models, depends just on the scale
$\Lambda$. Increasing $\Lambda$ requires a more effective
compensation, favoring greater values of $h$.

Apart of the possibility that $h$ is currently underestimated, let us
remind other substantial options which are still to be deepened: (i)
The contribution of topological singularities to DM were not
considered here and taking them into account could naturally increase
the amount of DM, yielding smaller $\Lambda$ values for the
double--axion model; fig. \ref{lambda} however shows that this is
hardly an efficient solution. (ii) The choice of a SUGRA potential is
however arbitrary, the dual axion model does not require SUGRA. Other
potentials can possibly yield the same coupling intensity in
agreement with a smaller $h$.

\section{Conclusions}
The first evidences of DM date some 70 years ago, although only in the
late Seventies limits on CMB anisotropies made evident that a
non--baryonic component had to be dominant. DE could also be dated
back to Einstein's {\it cosmological constant}, although only SNIa
data revived it, soon followed by data on CMB and deep galaxy samples.

Axions have been a good candidate for DM since early Eighties,
although various studies, as well as the occurrence of the SN 1987a,
strongly constrained the PQ scale around values $10^{10} \lesssim
F_{PQ} \lesssim 10^{12}$GeV. Contributions to DM from topological
singularities (cosmic string and walls) narrowed the constraints to
$F_{PQ}$.  Full agreement on the relevance of these contributions has
not yet been attained and, in this paper, they are still disregarded.

The fact that DM and DE can both arise from scalar fields, just by
changing the power of the field in effective potentials, already
stimulated the work of various authors. A potential like (\ref{eq:l1})
was considered in the so--called {\it spintessence} model
\citep{boyle02,gu01}. According to the choice of parameters, $\Phi$
was shown to behave either as DM or as DE. In the frame of a model of
tachyon DE, Padmanabhan \& Choudhury (2002) also built a model where
DM and DE arise from a single field. The possibility that both DM and
DE arise from the solution of the strong CP problem was also suggested
by Barr \& Seckel (2001). Their model, however, does not deal with
dynamical DE and aims to explain why the vacuum energy is so finely
tuned, while DM is simultaneously provided.

Here we deal with the possibility that the $\Phi$ field, which solves
the strong--$CP$ problem, simultaneously accounts for {\it both} DE
and DM.  The angle $\theta$ in eq.~(\ref{eq:n1}), as in the PQ model,
is turned into a dynamical variable, {\it i.e.} the phase of a scalar
field $\Phi$. While $\theta$ is gradually driven to approach zero, by
our cosmic epoch, in our model $\phi$ gradually increases and
approaches $m_p$, yielding DE. Residual $\theta$ oscillations,
yielding axions, account for DM.

The main topic of this paper is however the fit of this and other
models with WMAP data. We compared $\Lambda$CDM, SUGRA dynamical and
coupled DE models, as well as a scheme we dubbed $\phi^{-1}$--model
against these data. The last model encloses DM and DE and assumes that
they are coupled in a non--parametric way, with $C(\phi) = \phi^{-1}$.
This last model does not prescribe the origin of DE and DM, just as in
standard theories of coupled DE. The fit with WMAP constrains
$\Lambda$, and this constraints agree with the scale range required by
the dual--axion model; thus WMAP data support the dual--axion scheme.

The fits of WMAP data to $\Lambda$CDM, uncoupled and
constant--coupling SUGRA models, as well as to $\phi^{-1}$ SUGRA
models, yield similar $\chi^2$'s, for all models. At variance from
other model categories, however, in $\phi^{-1}$ models CMB data
constrain $\Lambda$. This is due to the stronger effects of $\phi$
variations on the detailed ISW effect, as they affect both DE pressure
and energy density, as well as DE--DM coupling.  In principle, this
strong impact of $\phi$ variation could badly disrupt the fit and make
$\phi^{-1}$ models significantly farther from data. This
does not occur, while the observational $\Lambda $ range agrees
with the dual--axion model at a 2--$\sigma$ level.

The success would be complete if the favored range of values of
the Hubble parameter ($h \sim 0.8$--1) could be slightly lowered.
This range is however obtained just for a SUGRA potential. This choice
is not compulsory and, moreover, contributions to axion DM due to
topological singularities were also disregarded. Furthermore, primeval
fluctuations were assumed to be strictly adiabatic while, in axion
models, a contribution from isocurvature modes can be expected. This
could legitimately affect the apparent position of the first peak in
the anisotropy spectrum, so completing the success of the model, in a
fully self--consistent way.

\end{document}